\documentclass[prb,preprint,amsmath,amssymb]{revtex4}
\begin{document}
\author{Kevin Leung}
\affiliation{Sandia National Laboratories, MS 1415,
Albuquerque, NM 87185\\
\tt kleung@sandia.gov}
\date{\today}
\title{First Principles Modeling of Mn(II) Migration above
and Dissolution from Li$_x$Mn$_2$O$_4$ (001) Surfaces}

\input epsf
 
\begin{abstract}

Density functional theory and {\it ab initio} molecular dynamics simulations
are applied to investigate the migration of Mn(II) ions to above-surface sites
on spinel Li$_x$Mn$_2$O$_4$ (001) surfaces, the subsequent Mn dissolution into
the organic liquid electrolyte, and the detrimental effects on graphite anode
solid electrolyte interphase (SEI) passivating films after Mn(II) ions diffuse
through the separator.  The dissolution mechanism proves complex; the
much-quoted Hunter disproportionation of Mn(III) to form Mn(II) is 
far from sufficient.  Key steps that facilitate Mn(II) loss include concerted
liquid/solid-state motions; proton-induced weakening of Mn-O bonds forming
mobile OH$^-$ surface groups; and chemical reactions of adsorbed
decomposed organic fragments.  Mn(II) lodged between the inorganic
Li$_2$CO$_3$ and organic lithium ethylene dicarbonate (LEDC) anode SEI
components facilitates electrochemical reduction and decomposition
of LEDC.  These findings help inform future design of protective coatings,
electrolytes, additives, and interfaces.

\vspace*{0.5in}
\noindent keywords: lithium ion batteries; lithium manganese oxide; solid
electrolyte interface; {\it ab initio} molecule dynamics;
computational electrochemistry

\end{abstract}

\maketitle
 
\section{Introduction}

Lithium ion batteries (LIB) featuring transition metal oxide cathodes
and organic solvent-based electrolytes are currently the energy storage
devices used to power electric vehicles.  Spinel lithium magnesium oxide
(Li$_x$Mn$_2$O$_4$ or ``LMO'')\cite{thackeray_rev} and nickel-doped high
voltage spinel (Li$_x$Ni$_{0.5}$Mn$_{1.5}$O$_4$, ``LNMO'' )\cite{ram_spinel,jow}
are promising cathode materials.  One impediment to their deployment is the
dissolution of transition metal ions, which can diffuse to the anode,
corrupt the solid electrolyte interphase (SEI) films protecting the
graphite anode, and accelerate battery capacity
fade.\cite{mnsei1,mnsei2,mnsei3,mnsei4,mnsei5,mnsei6,ram06} 
Such degradation is particularly severe at elevated temperature for
LNMO\cite{ram_spinel} and LMO.\cite{ram06,spinelht1,spinelht2,spinelht3}
Cation doping,\cite{tarascon91} surface coatings,\cite{xiao,jung15}
 and other means have been applied to reduce Mn dissolution,
but have so far not completely eliminated it.
Transition metal loss from other cathode materials is also widely
documented.\cite{ald0,nidiffus,ram06}   Understanding the detailed
Mn dissolution mechanism is crucial for new design principles that
can further mitigate this degradation route.

Mn dissolution has often been discussed in connection with Hunter's
disprortionation mechanism,
\begin{equation}
2 {\rm Mn(III)} \rightarrow {\rm Mn(II)} + {\rm Mn(IV)},
\end{equation}
associated with under-coordinated Mn(III) on LMO surfaces.\cite{hunter}
Early computational studies have focused on demonstrating the existence
of Mn(II) on LMO surfaces under ultrahigh vacuum (UHV) conditions.\cite{uhv}
A recent study, which includes explicit liquid solvent molecules,
reveals that the solvent can coordinate to surface Mn(III) ions, completing
their coordination shells, converting them to Mn(IV), and removing the
driving force for disproportionation.\cite{leung1}  More in-depth
atomic-lengthscale studies of Mn dissolution are clearly needed.

\begin{figure}
\centerline{\hbox{ \epsfxsize=4.00in \epsfbox{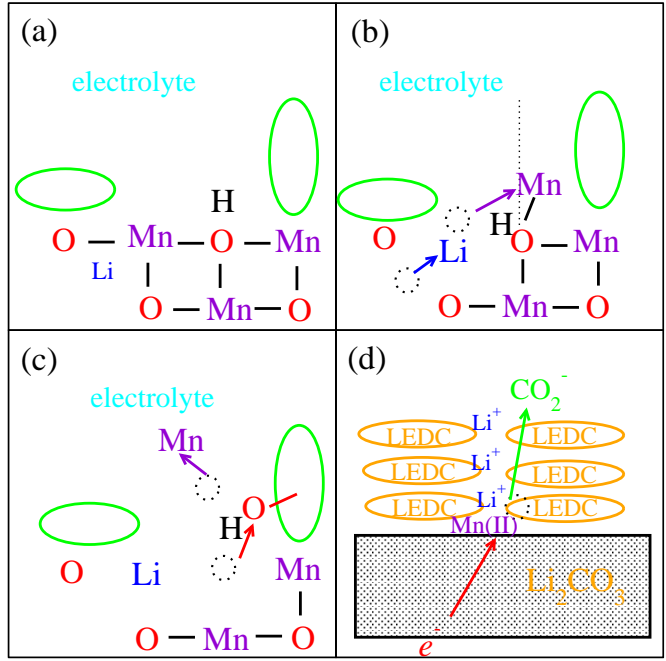} }}
\centerline{\hbox{ \epsfxsize=3.40in \epsfbox{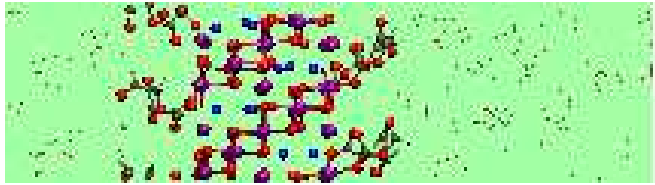} }}
\caption[]
{\label{fig1} \noindent
(a)-(d) Schematic depictions of key processes studied in this work.
(a) A layer of decomposed EC molecular fragments (green) on LMO (001);
(b) Mn(II) migration to above-surface sites;
(c) Mn(II) dissolution;
(d) Mn(II)-assisted reductive decomposition of LEDC to form CO$_2^-$.
(e) Snapshot of the AIMD simulation cell.
Li, Mn, F, O, C, and H atoms are in blue, purple, pink, red, grey, and
white, respectively.  The ``tagged'' Mn ion is green.  Atoms in
the oxide and decomposed EC molecule are depicted as spheres, while the
intact EC solvents are thin lines.
}
\end{figure}

Extensive spectroscopic and imaging studies have been conducted to understand
degradation on cathode oxide material surfaces.  The existence of liquid
electrolyte decomposition products, forming very thin SEI (sometimes called
cathode electrolyte interphase or CEI) films on the cathode,
has been amply demonstrated.\cite{kanno1,kanno2,ram_spinel,oh,song}
The overall speciation of electrolyte decomposition products have been
reported,\cite{edstrom,aurbach01,aurbach99} but elucidation of the atomic
structure and chemistry of the active material/SEI interface, expected to
be most relevant to Mn dissolution, remains a challenge.  The composition
of surface films is likely not static but depends on charge/discharge
conditions.\cite{novak,oh} Scanning transmission electron microscopy (STEM) and
other techniques have demonstrated that cycled layered nickel/manganese/cobalt
(NMC) oxides undergo surface reconstruction to a Mn(II) rock salt 
phase.\cite{doeff} Surface reconstruction and phase transformation have also
been reported for spinels at elevated temperatures in accelerated aging
studies.\cite{spinelht1,spinelht2,spinelht3} Under more standard battery
operation conditions and at shorter times, Li {\it et al}.  have applied
in-situ soft X-ray Absorption Spectroscopy (XPS) to reveal enhanced Mn(II)
content at LNMO/liquid electrolyte interfaces during battery charging, and
reduced Mn(II) surface concentration during discharge.\cite{wlyang} 
This is paradoxical, as more Mn(IV) are expected during the charging process.
They interpret the observation as evidence that Mn ions on the electrode
surface reacts with the liquid electrolyte.

Other studies have further emphasized the role of specific organic fragments
on Mn dissolution.  Jarry {\it et al.}\cite{jarry} have applied fluorescence
spectroscopy to identify $\beta$-diketonate chelating to dissolved Mn, and have
proposed a detailed mechanism for the formation of this species from the
dimethyl carbonate (DMC) cosolvent used in organic battery electrolytes.
Meanwhile computational studies have reported decomposition of ethylene
carbonnate (EC) on two LMO surface facets.\cite{leung1,leung2,borodin2015}
These studies report EC oxidation, ring opening, and proton transfer to LMO
surfaces.  They omit DMC and do not report $\beta$-diketonate formation.
However, it can be argued that Mn-chelating organic species other than
$\beta$-diketonate also facilitate Mn dissolution, even if they lack
fluoresence signatures and are not readily detected.  

These advances make it timely to revisit the mechanism associated
with Mn loss from spinel oxides.  We distinguish between two processes: Mn(II)
migration to non-crystallographic sites above the surface, and subsequent
Mn(II) dissolution, as shown schematically in Fig.~\ref{fig1}.  The
(001) surface of stoichiometric LiMn$_2$O$_4$ in vacuum exhibits
Li-sites half a lattice plane above surface oxygen ions.\cite{benedek_dry}
Hence it is not surprising that Mn(II) may occupy vacant surface Li-sites 
as dissolution intermediates.  Local minimum states associated with transition
metal ion desorption from mineral surfaces immersed in water have also
been reported.\cite{geochem}

We apply {\it ab initio} molecular dynamics (AIMD) simulations alongside
potential-of-mean-force (PMF) free energy techniques\cite{chandler} to
calculate the barriers asociated with these steps.  A precedent for this
study is Ref.~\onlinecite{benedek}, which focuses on LMO dissolution in an
aqueous, not organic, electrolyte.  That work predicts a dissolution timescale
which exceeds battery operation duration for 4-coordinated Mn(III) on the
(001) surface.  A major difference is that in water, auto-ionization of
H$_2$O molecules can release H$^+$ and OH$^-$ along the reaction pathway to
help break Mn-O bonds.  In organic solvents, that pathway is uavailable,
and the dissolution mechanism is expected to be more intricate.

We focus on the LMO (001) surface immersed in an EC liquid.  Mn ions are
exposed on clean (001) facets.  We decorate LMO (001) with partially decomposed
EC fragments and H$^+$ which are previously predicted,\cite{leung1} and find
that these species facilitate transition metal motion.  Note that
reconstructed (111)
is the most stable LMO facet.\cite{persson,liverpool,greeley,wolverton}
However, to model Mn dissolution from (111), it would first be necessary to
simulate the loss of surface O$^{2-}$, because all Mn ions reside in
subsurface layers.\cite{persson,liverpool,greeley,wolverton} For similar
reasons, we have not considered Li$_2$CO$_3$ films which have been reported on
cycled cathode surfaces.\cite{novak,aurbach99,co3,eriksson} Since there is
no empirical evidence that Mn(II) diffuse through the inorganic SEI components
like Li$_2$CO$_3$,\cite{shkrob,curtiss} the carbonate layer would have to
crack or be oxidatively destroyed during charging for Mn to
 dissolve.\cite{graindiffus,yue16} 
 Along these lines, atomic layer deposition
(ALD)\cite{xiao,jung15,ald0,ald1,ald2,ald3,ald4} and other surface protection
coatings\cite{mauger} have been applied to reduce Mn loss.  One way to
further improve ALD-coated electrode is to make them resistant to reactions
with the liquid electrolyte.  The extent to which organic electrolyte 
corrodes ALD protective layers can be examined with the
computational techniques applied in this work (see Sec.~\ref{discuss}).

In our modeling work, the exiting Mn(II) is coordinated to a F$^-$ anion,
which originates from decomposition of the standard battery electrolyte
counter-ion PF$_6^-$.  In the literature, hydrofluoric acid (HF), which 
arises from PF$_6^-$ reaction with trace H$_2$O in the liquid electrolyte,
has been strongly correlated with increased Mn(II)
dissolution.\cite{spinelht2,saulnier,sano,oh97}  In the supporting information
(S.I.) document, we show that HF is not needed to acidify LMO surface to yield
surface OH$^-$ groups that weaken transition metal-binding; trace water already
fills that function.  Hence it is reasonable to speculate that a main role of
HF is to provide F$^-$ that binds to Mn, at least during the exit from
the cathode.\cite{borodin_h+}  Some evidence of MnF$_2$ in the anode SEI has
been reported.\cite{xiao14}  However, MnF$_2$ XPS signatures are similar to
those of LiF, which complicates its detection on cathode
surfaces.\cite{eriksson,oh12}

Finally, we address one possible way Mn(II) ions which have diffused through
the separator into anode SEI films may disrupt SEI functions.\cite{mnsei6}
The structure, dispersion, and location of Mn inside the anode is controversial
(see Shkrob {\it et al.}\cite{shkrob} for a critical overview); here we
only focus on its catalytic function.  Two recent studies\cite{shkrob,curtiss}
have concluded that Mn exist as isolated Mn(II) complexes situated between
inorganic (e.g., Li$_2$O, LiF, but especially\cite{shkrob} Li$_2$CO$_3$) and
organic (e.g., lithium ethylene dicarbonate, LEDC) SEI components.  It is
proposed that solvent molecules can diffuse through the porous organic region
to coordinate to Mn(II) ions and electrochemically reduce them.\cite{curtiss}
Mn(II) thus act as conduits for $e^-$ transfer to solvent molecules.  However,
this scenario requires pores in the SEI large enough to transmit solvent
molecules, which have not been reported in classical force field molecular
dynamics simulations.\cite{borodin}  Here we show that Mn(II) plus excess
electrons can instead decompose LEDC in redox reactions (Fig.~\ref{fig1}d).
This mechanism dovetails with recent computational demonstrations that SEI
components are not as chemically stable as generally believed.\cite{batt,perla}
Other experimental studies have suggested the existence of transition metal
particles inside the anode SEI.\cite{xiao14,nidiffus,ochida,gowda}  Although
this is disputed by electron paramagnetic spectroscopy,\cite{shkrob}  
we also report simulations of LEDC decomposition on Ni(111) surfaces in the
S.I.~to address this possibility.

This manuscript is organized as follows.  Sec.~\ref{method} describes the
methods used and includes a discussion on the challenges of modeling
solid-liquid interfaces.\cite{interface}  Sec.~\ref{results} discusses the
results.  Sec.~\ref{discuss} extrapolates the predictions to design principles 
or protective coatings.  Sec.~\ref{conclusion} concludes the paper with a
summary of the main findings.

\section{Methods}
\label{method}

\subsection{Computational Challenges and Perspectives}

First we briefly discuss the limitations of AIMD modeling
of buried liquid-solid interfaces.  Imaging and spectroscopic techniques like
TEM and XPS have yet to provide atomic lengthscale-resolution structural
data that should be the starting points of such calculations.  In particular,
the precise surface features exposed at the interface, the speciation of
decomposed electrolyte fragments (``SEI'') adsorbed on the cathode,
the thickness of such SEI layers, and the identity and surface concentration
of structural defects that may enhance Mn dissolution, have not been
elucidated.  In electrochemical settings, there is the additional challenge
of determining the surface charge density consistent with the applied voltage.  

Our approach is to create model surface structures and study their most
relevant properties.  In this case, the key metrics are Mn surface migration
and dissolution barriers.  If the predicted time frames exhibit large
discrepancy with measurments, the models are modified.  Thus our
interfacial model structures should be considered plausible scenarios.
Even with this caveat, these calculations are valuable for providing insights
into battery degradation at the atom-by-atom lengthscale and bond-breaking 
sequence-of-event detail not yet available to measurements.  These insights
will help guide future design of cathode protective coatings and strategies.
This work sidesteps the issue of the applied voltage.  LMO is a small-polaron
conductor, and the Fermi level of the metallic current-conductor attached
to it should be in equilibrium with polaron formation energies at the
solid electrode/liquid electrolyte interface.\cite{leung3}

\subsection{Simulation Details}

Finite temperature {\it ab initio} molecular dynamics (AIMD) simulations
are conducted under solvent-immersed conditions.  A few static ultrahigh
vacuum (UHV) DFT calculations, performed at T=0~K, are also reported.  All
calculations apply periodical replicated simulation cells, the Vienna Atomic
Simulation Package (VASP) version 5.3,\cite{vasp1,vasp1a,vasp2,vasp3} and the
Perdew-Burke-Ernzerhof (PBE) functional.\cite{pbe}  Modeling spinel
Li$_x$Mn$_2$O$_4$ requires spin-polarized DFT+U augmented treatment of Mn $3d$ 
orbitals.\cite{dftu1} The $U$ and $J$ values depend on the orbital projection
scheme and DFT+U implementation details; here $U-J=$4.85~eV is chosen in
accordance with the literature.\cite{zhou,liverpool1} A 400~eV planewave energy
cutoff and $\Gamma$-point Brillouin zone sampling are imposed.  The bulk
LiMn$_2$O$_4$ crystal is antiferromagnetic (AFM).  We have imposed an AFM
ordering on alternate Mn planes along the (011) direction.  The electron spin
on each Mn ion is examined to determine Mn charge states.

Static geometry optimization simulation cells are of dimensions
34$\times$11.88$\times$11.88\AA$^3$.  They have a Li$_6$Mn$_{20}$O$_{32}$H$_8$
cathode oxide stoichiometry plus eight EC fragments (C$_3$O$_4$H$_3$).  
A 10$^{-4}$~eV convergence criterion is imposed.  

The AIMD simulation cell is similar to the static cells, but the 
$z$-dimension perpendicular to the interface is increased to 46.5~\AA$^3$.
32 EC molecules are confined in the space between the organic
fragment-decorated oxide surfaces.  The cell is pre-equilibrated using simple
molecular force field and Monte Carlo (MC) simulations.\cite{ec,leung1}.
AIMD simulations start from the final MC-generated configuration.  A
10$^{-6}$~eV convergence criterion is imposed at each AIMD
Born-Oppenheimer time step.  The trajectories are kept at an average
temperature of T=450~K using Nose thermostats.  Tritium masses on EC are
substituted for proton masses to permit a time step of 1~fs.  Under these
conditions, AIMD trajectories exhibit drifts of less than 2~K/ps.  The
coordinates and velocities obtained from a 11.5~ps AIMD equilibration
trajectory are used in potential-of-mean-force (PMF or $\Delta W(R)$)
calculations.  

All PMF simulations apply two-body reaction coordinates of the form
$R$=$|{\bf R}_{\rm Mn}-{\bf R}_{\rm O}|$ or $R'$=$z_{\rm Mn}$-$z_{\rm Mn'}$,
where ${\bf R}$ is the position vector of an atom and $z$ is its position
perpendicular to the interface.  The two coordinates are used in Mn surface
migration (Sec.~\ref{migrate}) and dissolution (Sec.~\ref{dissolve}) studies,
respectively.  The specific atoms involved will be described in
Sec.~\ref{results}.  Harmonic penalties $B_o(R-R_o)^2/2$ or $B_o(R'-R_o)^2/2$ 
are added to DFT+U energies in a series of windows with a progression of
$R_o$ values, separated by 0.3~\AA, spanning the reaction paths. $B_o$ is set
at 4~eV/\AA$^2$.  $\Delta W(R)=-k_{\rm B}T \log P(R)$ where $P(R)$ is the
probability that a $R$ value is observed, after adjustment to remove the
effect of the umbrella sampling constraint.\cite{chandler}  A similar 
procedure is used for the coordinate $R'$.  The elevated temperature is only
adopted to accelerate the molecular dynamics.  The final $\Delta W(R)$
and $\Delta W(R')$ expressions assume a temperature of T=300~K.

Along the $R'$-coordinate, each window is initiated by taking a
configuration 1~ps into the trajectory from the previous window.  For
the $R$-coordinate, a new window is generated a few~ps into the trajectory
from three windows away ($\Delta R_o$=-0.9~\AA).  These tentpole windows are
then used to create starting configurations in adjacent windows with
$\Delta R_o$=$\pm$0.3~\AA.  This scheme is adopted to pre-estimate
the size of the barrier before computing the entire $\Delta W(R)$.
The first 1~ps in each window is used for equilibration and discarded. 
Statistics are collected for the next 10~ps.  Statistical uncertainties
in $W(R)$ are estimated by splitting the trajectory in each window into
five, calculating the standard deviation, and propagating the
noise across windows assuming gaussian statistics.  

We do not apply the popular metadyamics method, based on non-equilibrium
trajectories, to compute the PMF.\cite{meta} The main reason is that Mn
migration involves many moving parts, and diffusive molecular motions are
critical.  The umbrella sampling approach used herein permits us to run
trajectories of variable lengths that are not determined ahead of time.
This allows a more systematic treatment of diffusive motion.

In the absence of F$^-$ ions, our PMF calculation yields a barrier that
is too high compared to experimental timescales.  See the S.I.  To obtain
a lower barrier, we have manually added a F$^-$ anion coordinated to the
migrating or ``tagged'' Mn.  A further 1.4~ps AIMD simulation is conducted
for equilibration purpose, and a PMF calculation is restarted with this added
F$^-$.   

A different set of AIMD simulations involve the interface between LEDC and
Li$_2$CO$_3$ (Fig.~\ref{fig1}d).  
The cell size is 16.59$\times$19.79$\times$30~\AA$^3$.
The lateral cell dimensions are those of the Li$_2$CO$_3$ (001) surface
cell.  A bilayer of LEDC, which exhibits molecular ``crystal structure'' with
slightly smaller lattice constants,\cite{batt} is placed on the lithium
carbonate surface, and the atomic coordinates are optimized.  Unlike
cathode simulations, modeling of the LEDC/Li$_2$CO$_3$ interface includes a
vacuum region in the simulation cell.

Finally, some limited thermodynamics calculations are conducted using the
DFT/PBE0 functional.\cite{pbe0}

\section{Results}
\label{results}

\subsection{Ultrahigh Vacuum Condition Calculations}
\label{uhv}

\begin{figure}
\centerline{\hbox{ (a) \epsfxsize=2.00in \epsfbox{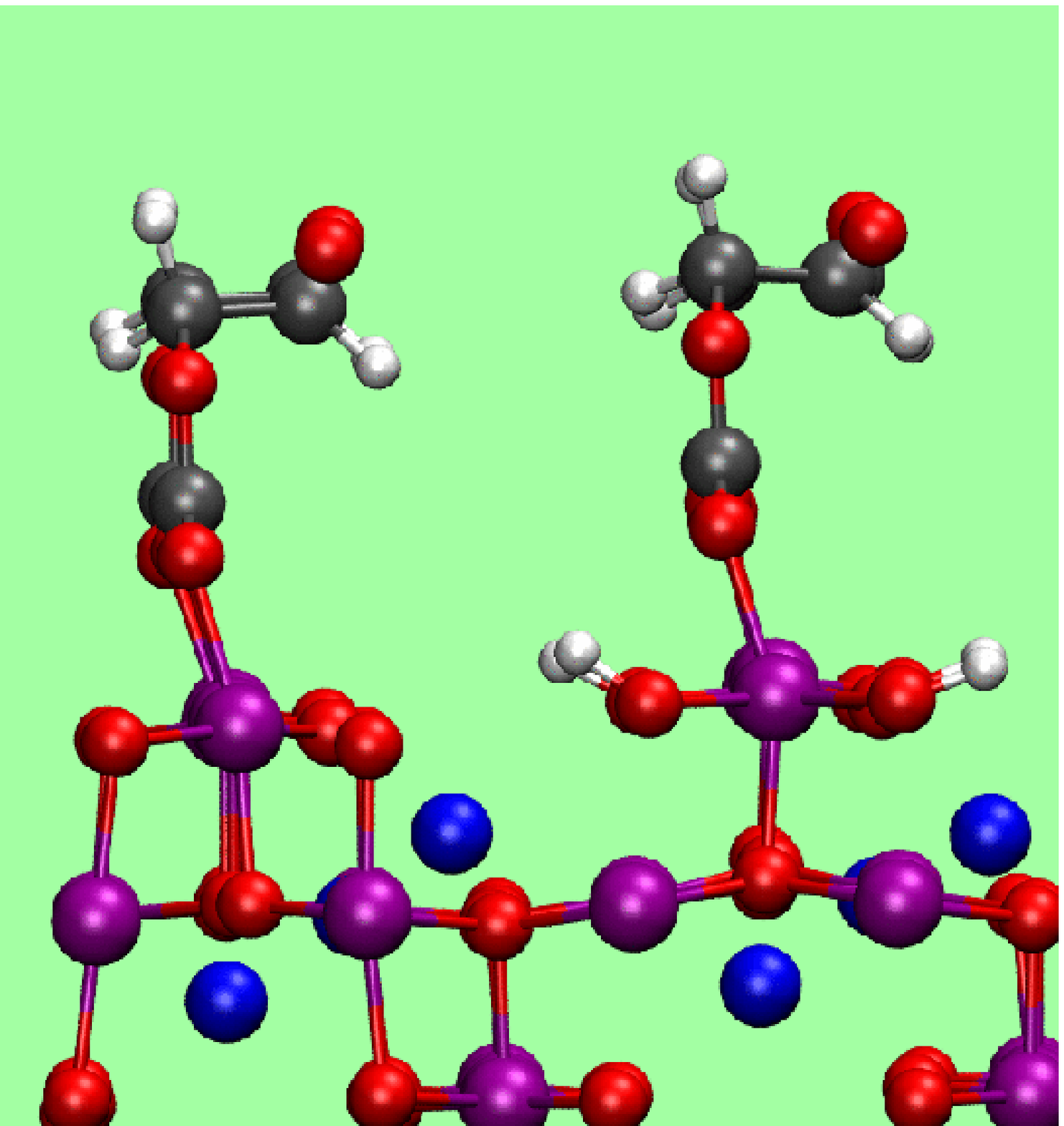}  
                   \epsfxsize=2.00in \epsfbox{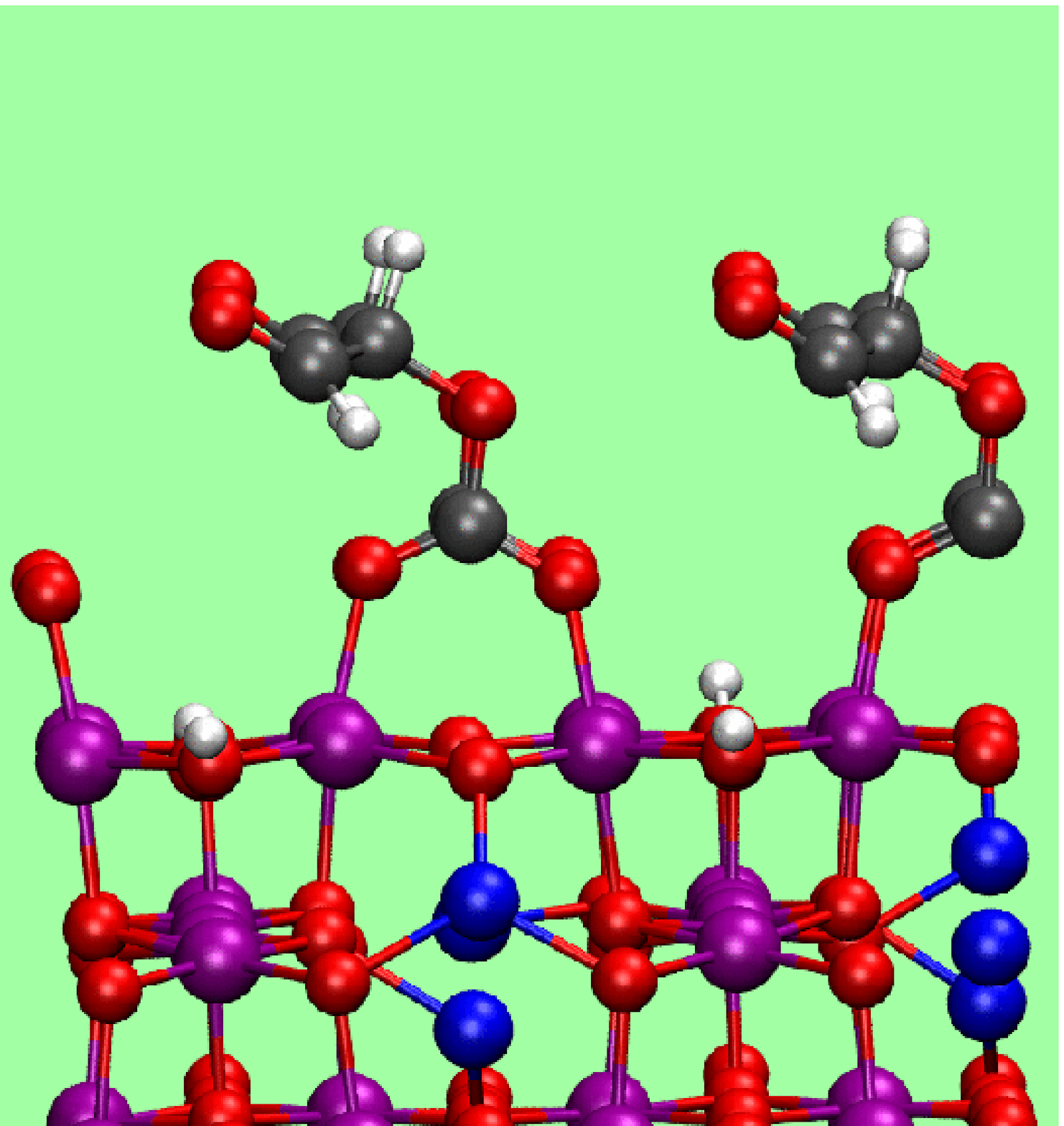} (b) }}
\centerline{\hbox{ (c) \epsfxsize=2.00in \epsfbox{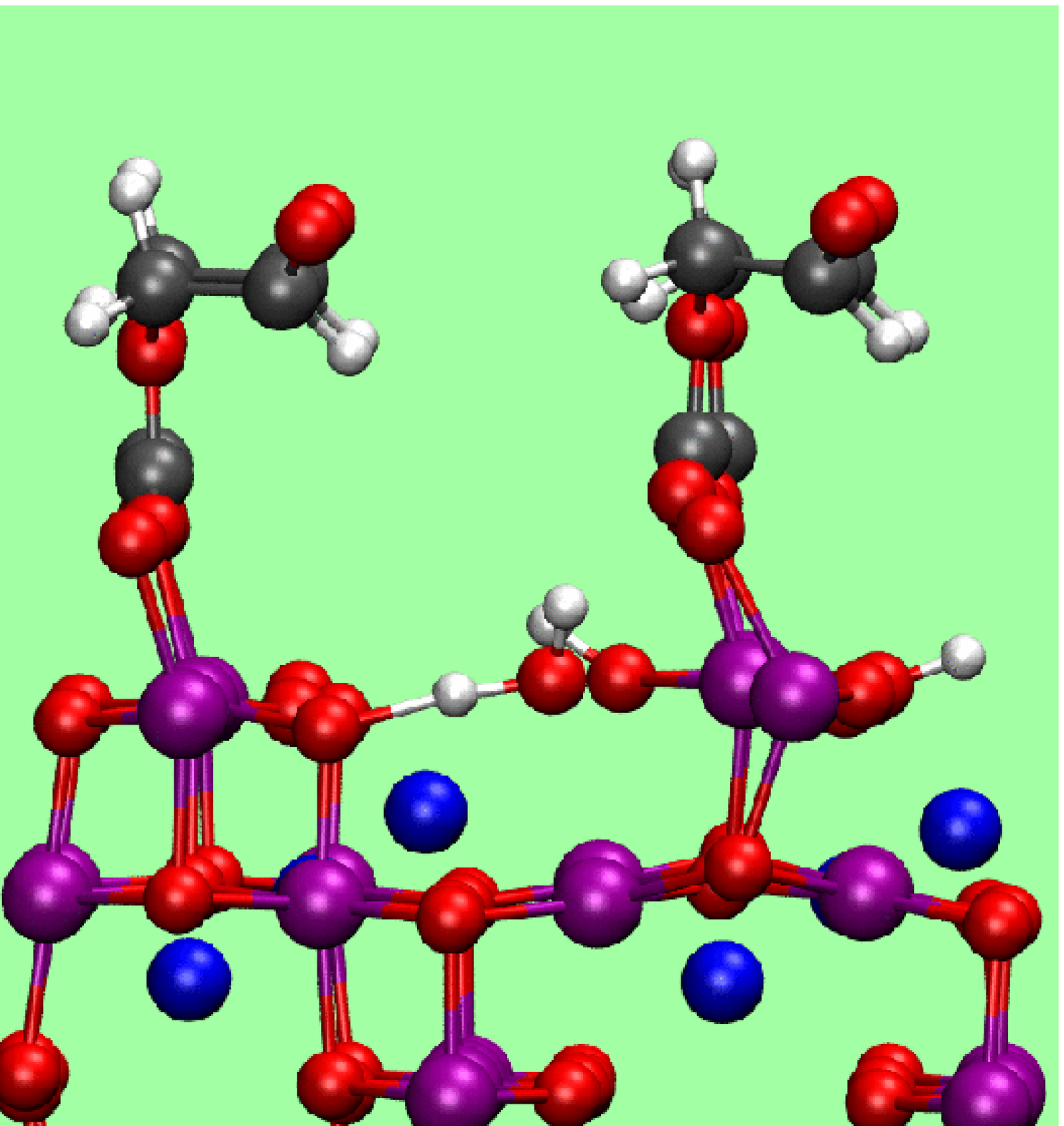} }}
\caption[]
{\label{fig2} \noindent
(a)-(b) Two perspectives of the 4 decomposed EC molecules covering each surface
of the periodically replicated, dry LMO (001) simulation cell; (c) attempt
to create a H$_2$O molecule.  The color key is the same as in Fig.~\ref{fig1}e.
}
\end{figure}

Figs.~\ref{fig2}a-b depict two perspectives of the dry LMO (001) surface
optimized at T=0~K.  Each of the four adsorbed, partially decomposed EC
fragments coordinates to two surface Mn ions via two of its three CO$_3^-$
oxygens.  Thus initially, each surface Mn ion is coordinated to four LMO
framework O$^{2-}$ and/or OH$^-$ plus a O atom of the EC fragment.  It would
have been 6-coordinated except for the surface O-vacancies, created when each
EC fragment abstracts one O$^{2-}$ from the (001) surface and donates a
H$^+$ to form Mn-OH-Mn bridges.\cite{leung1,leung2}  The multi-proton
configuration depicted in Fig.~\ref{fig2}a-b optimizes the energy.  This is
the starting point of liquid-state Monte Carlo pre-equilibration of liquid EC
configuration, during which the LMO and decomposed EC fragment atoms are frozen.

Fig.~\ref{fig2}c examines the possibility that two Mn-OH groups may 
disproportionate into Mn-O-Mn and H$_2$O.  The reaction is endothermic
by 1.21~eV.  While the removal of a surface O$^{2-}$ by two H$^+$ would have 
yielded Mn ions which are even more under-coordinated, and facilitated their
dissolution, no justification for H$_2$O formation is found.  This behavior
is in contrast to the clean LMO (001) surface saturated with
OH$^-$ groups.\cite{leung1}

\subsection{Migration of Mn to an Above-Surface Position}
\label{migrate}

\begin{figure}
\centerline{\hbox{ (a) \epsfxsize=2.00in \epsfbox{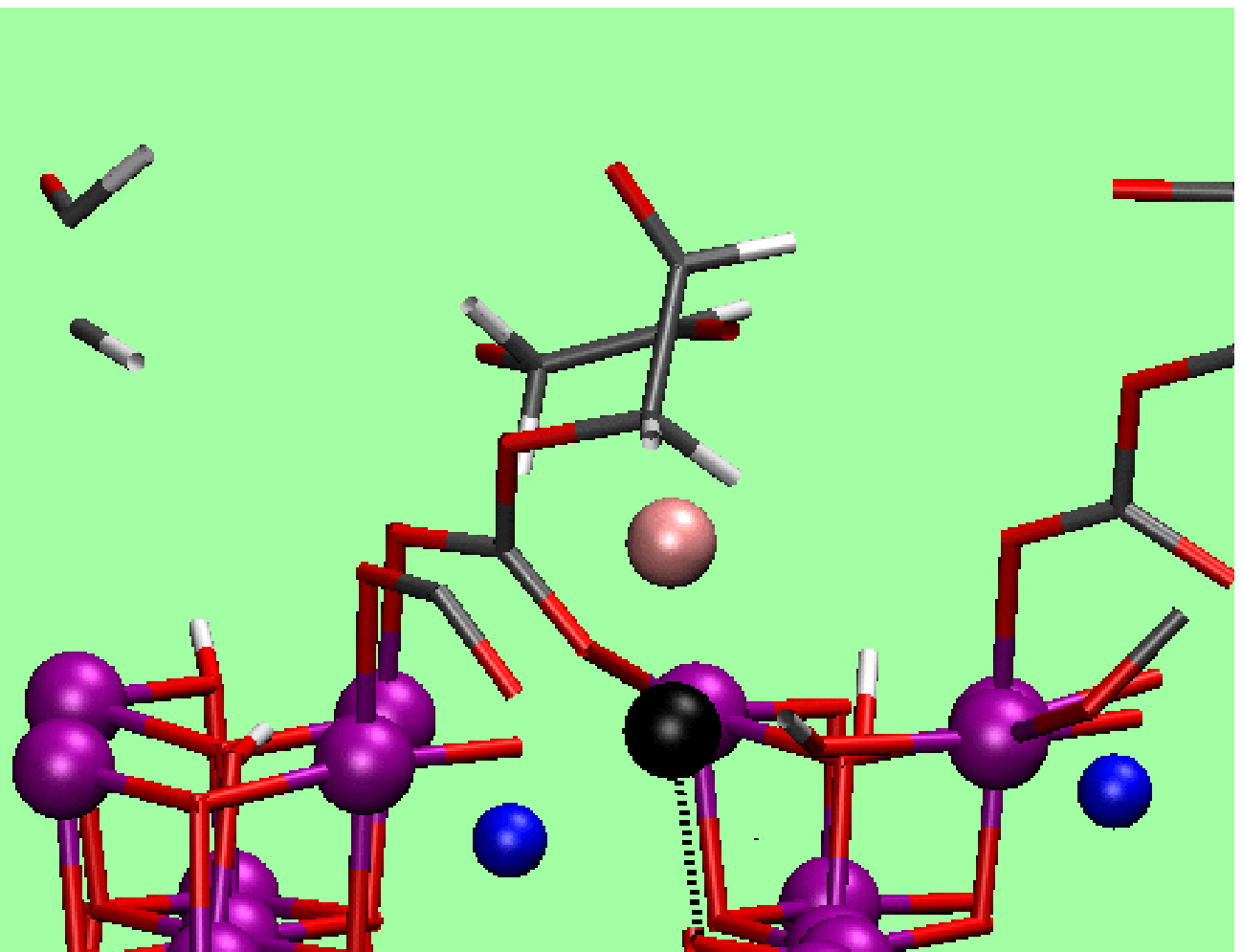}  
                   \epsfxsize=2.00in \epsfbox{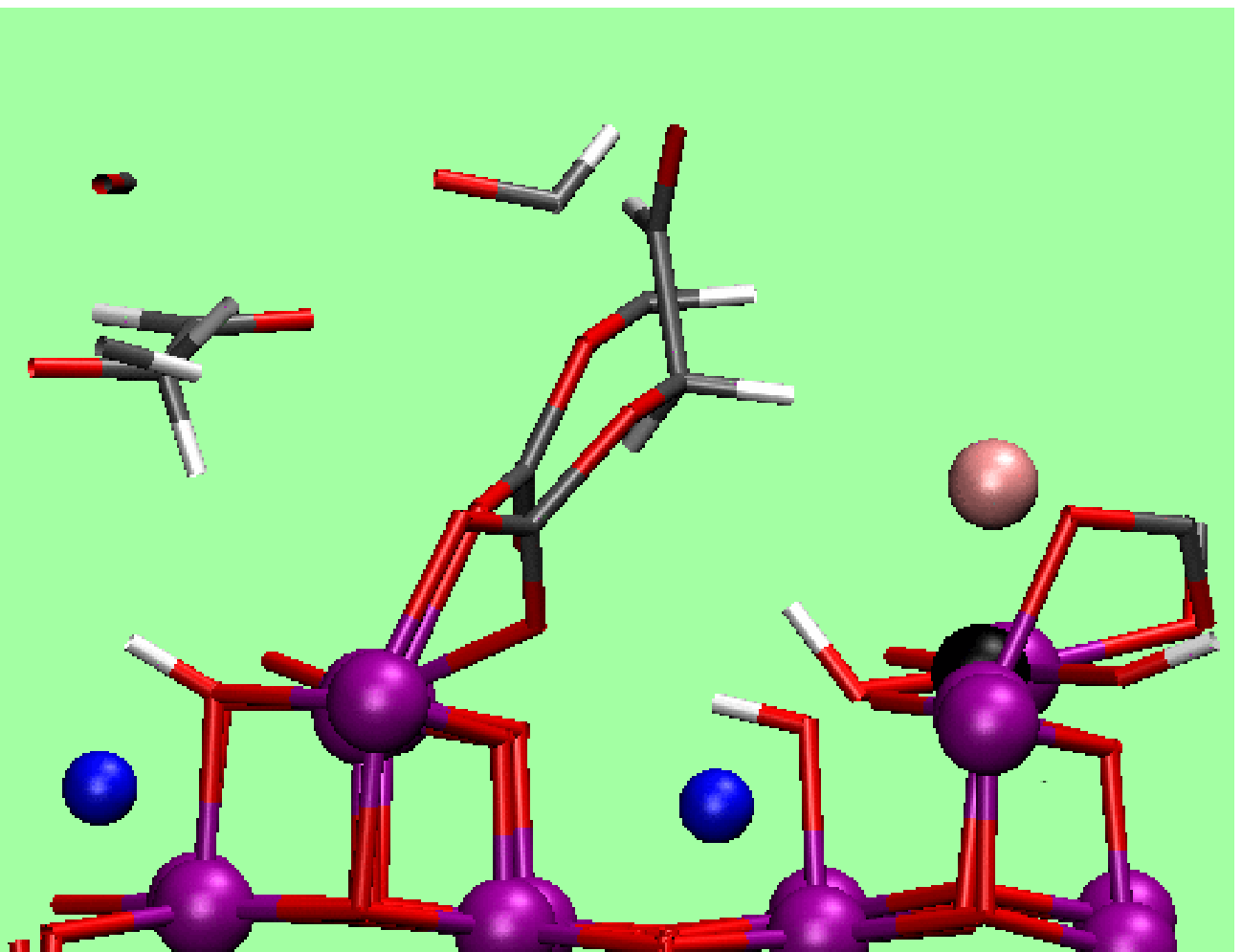} (b)}}
\centerline{\hbox{ (c) \epsfxsize=2.00in \epsfbox{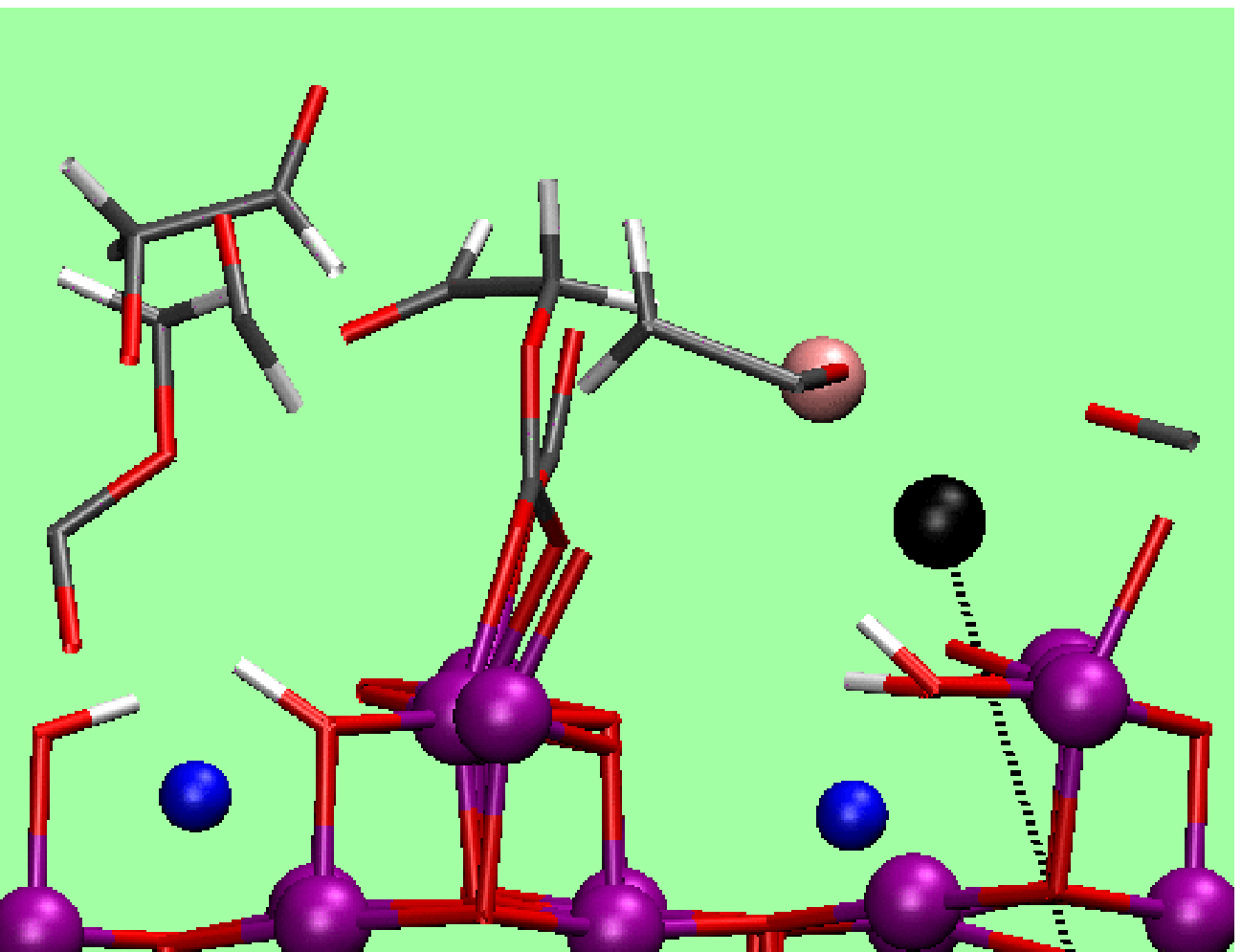}  
                   \epsfxsize=2.00in \epsfbox{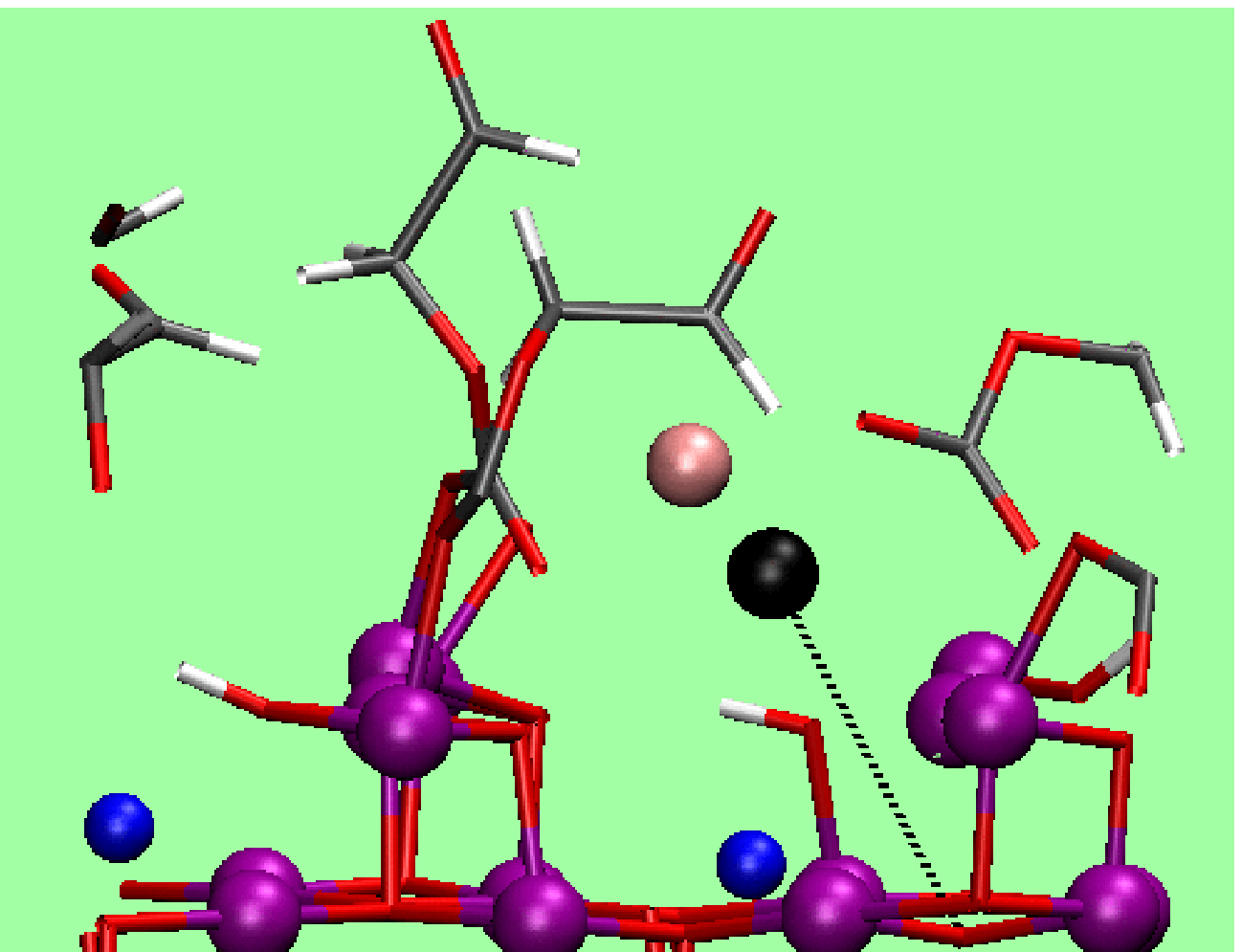} (d)}}
\centerline{\hbox{ (e) \epsfxsize=2.00in \epsfbox{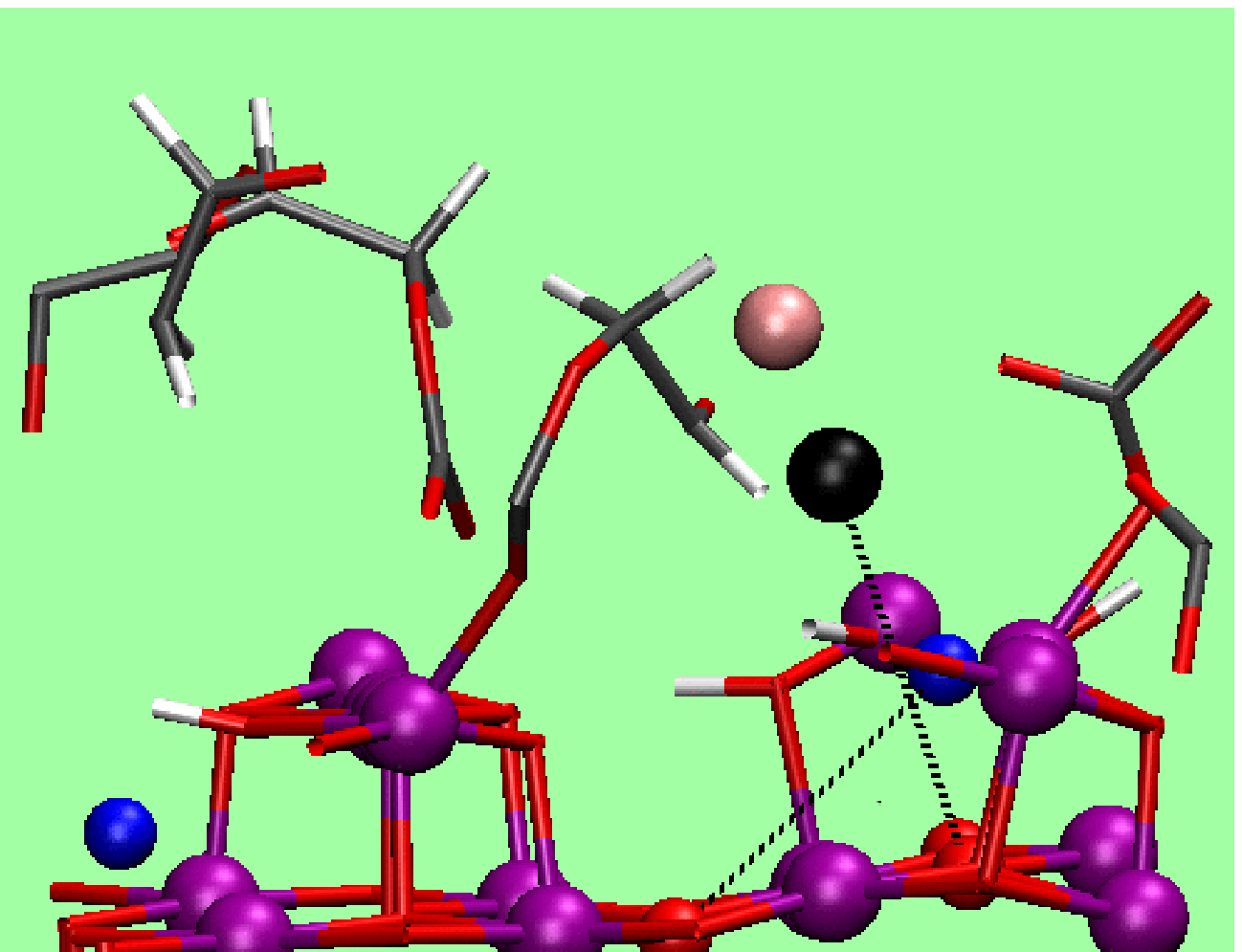}  
                   \epsfxsize=2.00in \epsfbox{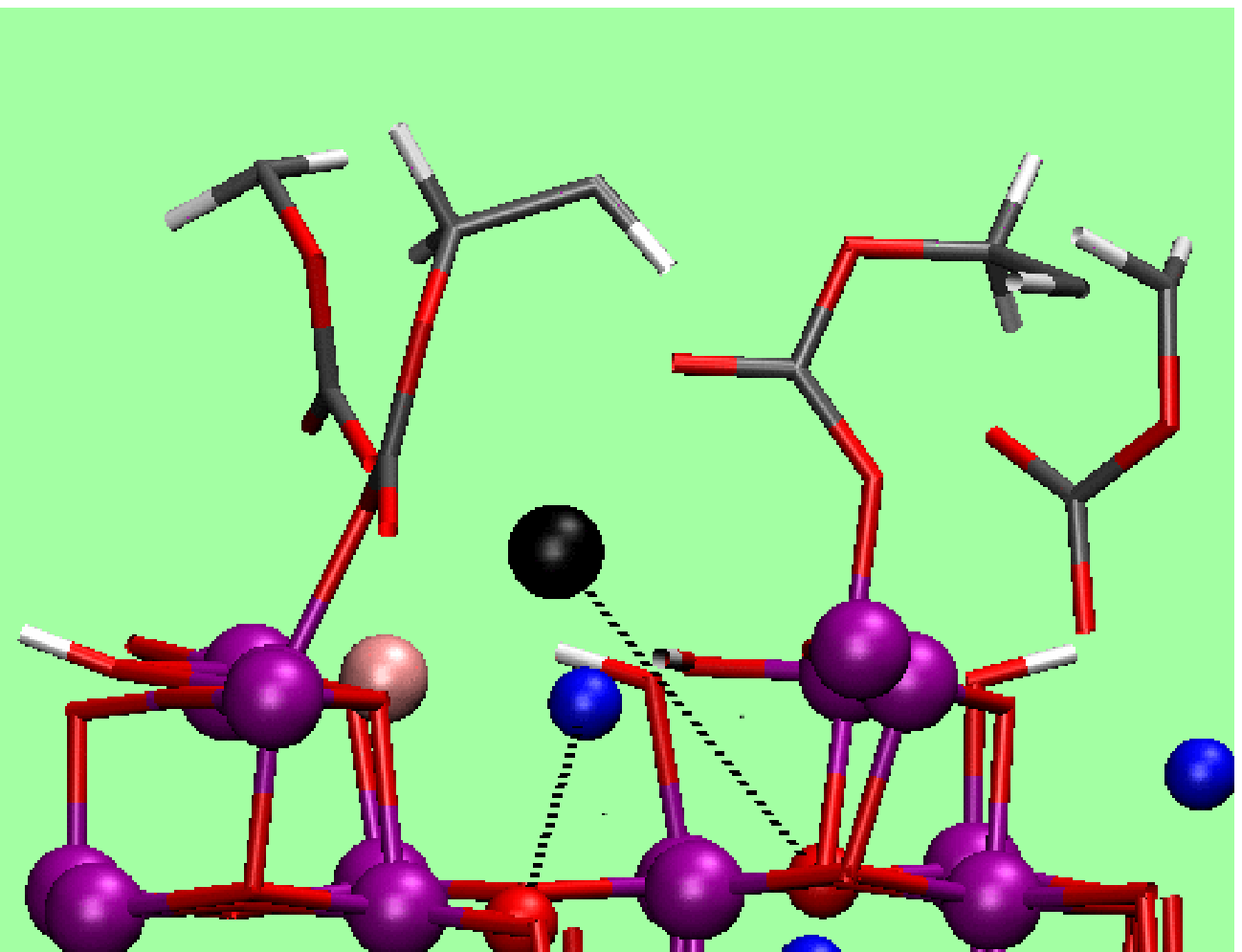} (f)}}
\caption[]
{\label{fig3} \noindent
(a)-(f) Snapshots along the reaction profile of Fig.~\ref{fig4}.
Panel (a) depicts the same configuration as (b) with a 90$^o$
rotation.  The color scheme is as in Fig.~\ref{fig2}.  O and H atoms
are depicted as stick figures; only Mn, Li, and F ions are spheres, and
intact solvent molecules are removed for clarity.  The tagged Mn ion
is in black.  The two blue dashed lines represent Mn(II)-O (``$R$'') and
Li$^+$-O distances (see text).
}
\end{figure}

During AIMD equilibration (before any PMF harmonic constraint is imposed), a
few of the decomposed EC fragments have one of their O atoms collapse on to
LMO surface oxygen vacancy sites, leaving some Mn uncovered.  This allows
those Mn to be coordinated to the electrolytes and actually facilitates
Mn removal.  Mn must ultimately bind to electrolyte molecules (EC and/or
anions) to dissolve.  During Mn departure from the surface, the EC fragment
O atoms weakly-bound to LMO can be readily pulled off the surface again.  As
discussed in Sec.~\ref{method}, we tag one such exposed Mn as the dissolving
species, add a F$^-$ to it, and use the harmonic constraints inherent to
$\Delta W(R)$ simulations to progressively pull it off its initial lattice
site.  See Fig.~\ref{fig3}a, which also depicts the reaction coordinate $R$.

\begin{figure}
\centerline{\hbox{ \epsfxsize=4.50in \epsfbox{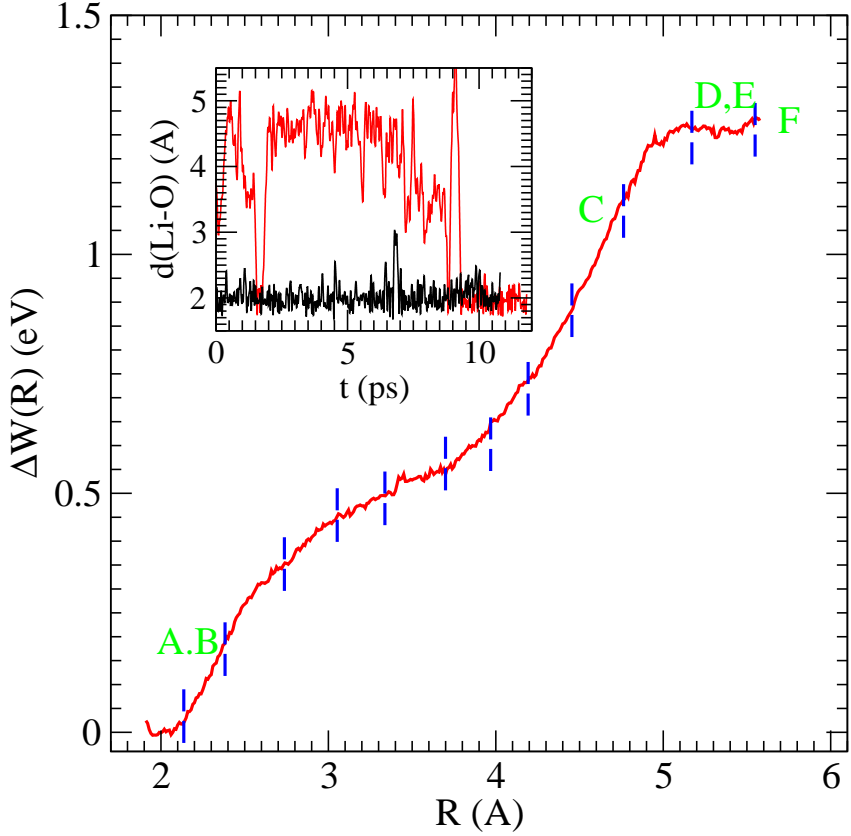} }}
\caption[]
{\label{fig4} \noindent
Free energy profile of Mn migration to above the surface.  ``A''-``F''
correspond to the panels in Fig.~\ref{fig3}a-f.  Inset: distance between a
subsurface Li$^+$ and a subsurface O$^{2-}$ ion initially coordinated to it.
The black and red lines are for sampling windows C and D/E, respectively.
}
\end{figure}

The $\Delta W(R)$ associated with the above-surface migration of this Mn is
shown in Fig.~\ref{fig4}.  Fig.~\ref{fig3}c depicts the tagged Mn ion near
the onset of the $\Delta W(R)$ plateau.  It is now 4-coordinated: to the
F$^-$, a decomposed EC fragment, and a O$^{2-}$ and a OH$^-$ on the LMO
surface.  The carbonyl oxygen of an intact EC molecule that has diffused
to the vicinity of the Mn is only slightly further away, not shown in this
snapshot; it binds to Mn in part of the trajectory.  Fig.~\ref{fig3}d
depicts a configuration taken from the plateau window.  The Mn ion has
``rolled over'' to the other side of the axis formed by the surface
O$^{2-}$ and OH$^-$ (see Fig.~\ref{fig1}b for illustration).  

A harmonic constraint is not needed or used in the plateau sampling window
(Fig.~\ref{fig3}d-e); the system is metastable in this region for at least
10~ps, showing that it is a reaction intermediate.  The onset of the plateau 
coincides with the activation of concerted solid- and liquid-state motions.
At this point, the tagged Mn has become sufficiently far from the LMO
surface that a subsurface
Li$^+$ ion can reversibly occupy the site vacated by this Mn (Fig.~\ref{fig3}e).
The time-dependence of this motion is illustrated by the red line in the inset
of Fig.~\ref{fig4}.  The distance between this Li$^+$ and a subsurface
O$^{2-}$ to which it is initially coordinated fluctuates between 2 and
4.5~\AA\, within picosecond time scales as the Li$^+$ moves back and forth.
For a comparison, the black line corresponds to the pre-plateau sampling
window ``C'' (Fig.~\ref{fig3}c), where the tagged Mn ion still repels Li$^+$
intrusion.  The corresponding Li$^+$/O$^{2-}$ distance fluctuates around
$\sim$2~\AA, the typical length of a stable Li$^+$/O$^{2-}$ ionic bond.   

Before PMF constraints are applied, all Mn on the surface exhibit +3 charge
states in every snapshot we have examined.  Thus no stable Mn(II) is found on
the liquid-immersed, EC fragment-decorated LMO (001) surface.  In the absence
of the F$^-$ (S.I.), when the Mn-O$^{2-}$ distance $R$ is 2.5~\AA\, or beyond,
the tagged cation is found to have gained an $e^-$ from other Mn(III) to
become a Mn(II) in all AIMD snapshots examined.  Fig.~S1 shows that such
a displacement costs less than 0.25~eV.  At this point in the reaction
profile, the system is at least $\sim$1.2~eV from reaching the end-of-reaction
point (S.I.).  This emphasizes that, in terms of dissolution kinetics on this
surface, Hunter disproportionation\cite{hunter} is far from sufficient.
When F$^-$ is present in the simulation cell (Figs.~\ref{fig3}-\ref{fig4}),
the tagged Mn can fluctuate between +2 and +3 charge states in AIMD snapshots
taken about 0.7~ps apart.  In the $R$$\sim$3~\AA\, and plateau windows in
Fig.~\ref{fig4}, this Mn is a Mn(II) in 53\% and 72\% of the snapshots,
respectively.  

Comparing $\Delta W(R)$ at $R$=2.0~\AA\, and 5.5~\AA, the free energy barrier
$\Delta G^*$ is determined to be 1.25$\pm$0.09~eV.  The uncertainty corresponds
to twice the standard deviation.  The predicted $\Delta G^*$ is consistent with
a reaction rate of one every 27000 years at room temperature if a canonical
kinetic prefactor of 10$^{12}$/s is assumed.  Experimentally, Mn dissolution is
observed within a much shorter time frame.\cite{mnsei1} Therefore $\Delta G^*$
is overestimated.  A 1.05~eV barrier would correspond to a more reasonable
117-hour rate at room temperature.  For a comparison, the dissolution barrier
of 4-coordinated Mn(III) at the LMO (001)/water interface has been predicted
to be either 1.8 or 1.4~eV, depending on whether a rescaling factor is
used.\cite{benedek} These values correspond to Mn ejection into liquid water
without going through an intermediate.  In principle, the ``coordination
number'' reaction coordinate used in Ref.~\onlinecite{benedek}
can support a two-step mechanism where Mn settles into an above-surface,
non-crystallographic intermediate,\cite{geochem} but the free
energy valley associated with this may be too small to detect.

Imposing harmonic biases with larger $R_o$ does not lift the tagged Mn ion
into the liquid electrolyte because the reaction coordinate $R$ allows
the Mn to slide along the surface instead.  Imposing a constraint on the
Li$^+$-subsurface O$^{2-}$ distance (inset of Fig.~\ref{fig4}), only, leads
to larger $R$ values, but the F$^-$ anion now falls on to the oxide surface
(Fig.~\ref{fig3}f).  Since F$^-$ is not part of any reaction coordinate,
this event is irreversible and prevents further free energy calculations.

The existence of Mn(II) above the surface oxygen plane, coordinated to
decomposed organic fragments, appears consistent with the interpretation of
Ref.~\onlinecite{wlyang}.  This experimental work emphasizes enhancement of
Mn(II) concentration at the liquid/solid interface during battery charging,
where the cathode voltage is high and more electrolyte decomposition should
occur.  One caveat is that Fig.~\ref{fig4} seems to suggest that the free energy
change ($\Delta G$) of Mn surface migration is identical to the barrier value
($\Delta G^*$).  If so, $\Delta G$=1.25~eV would imply an extremely small
population of Mn(II) on the surface.  This argument ignores further steps,
beyond the scope of our calculations, that can stabilize surface Mn(II).  One
is the continuous diffusion of Li$^+$ ions from the bulk oxide region to the
surface, occupying the original Mn surface site, blocking Mn return, and
yielding a favorable entropy change.  This is especially likely during charging
when Li$^+$ are being de-intercalated.  The concerted diffusion of a chain
of Li$^+$, reminiscent of the ``knock-on'' effect in Li$_2$CO$_3$,\cite{yueqi}
is at present beyond AIMD studies.  This is because Li$^+$ bulk diffusion 
exhibits barriers of 0.2-0.6~eV inside LMO.\cite{sd1,sd2,sd3}  Such barriers,
although not excessive, correspond to Li$^+$ motion time-scales that exceed
AIMD trajectory lengths, unless the non-equilibrium metadynamics
technique\cite{meta} is used to deal with all possible Li$^+$ diffusion
degrees of freedom.  Another Mn stabilization scenario is the diffusion of EC
molecules or surface organic fragments to complete the octahedral Mn(II)
solvation shell.  The tagged Mn is undercoordinated in Fig.~\ref{fig3}e.
Steric hindrance and slow EC diffusion have likely prevented it from reaching
6-coordination so far in the trajectory.  Note that Ref.~\onlinecite{wlyang}
involves LNMO while our model is Ni-free; therefore quantitative agreement
should not be expected.

\subsection{Mn(II) Dissolution from the LMO Surface}
\label{dissolve}

This subsection describes the dissolution of the above-surface Mn ion into
the liquid electrolyte.  For this purpose, we assume this configuration is
at zero free energy, due to stabilization events not included in
Sec.~\ref{migrate}, and ignore the work done to arrive at it.

\begin{figure}
\centerline{\hbox{ (a) \epsfxsize=2.00in \epsfbox{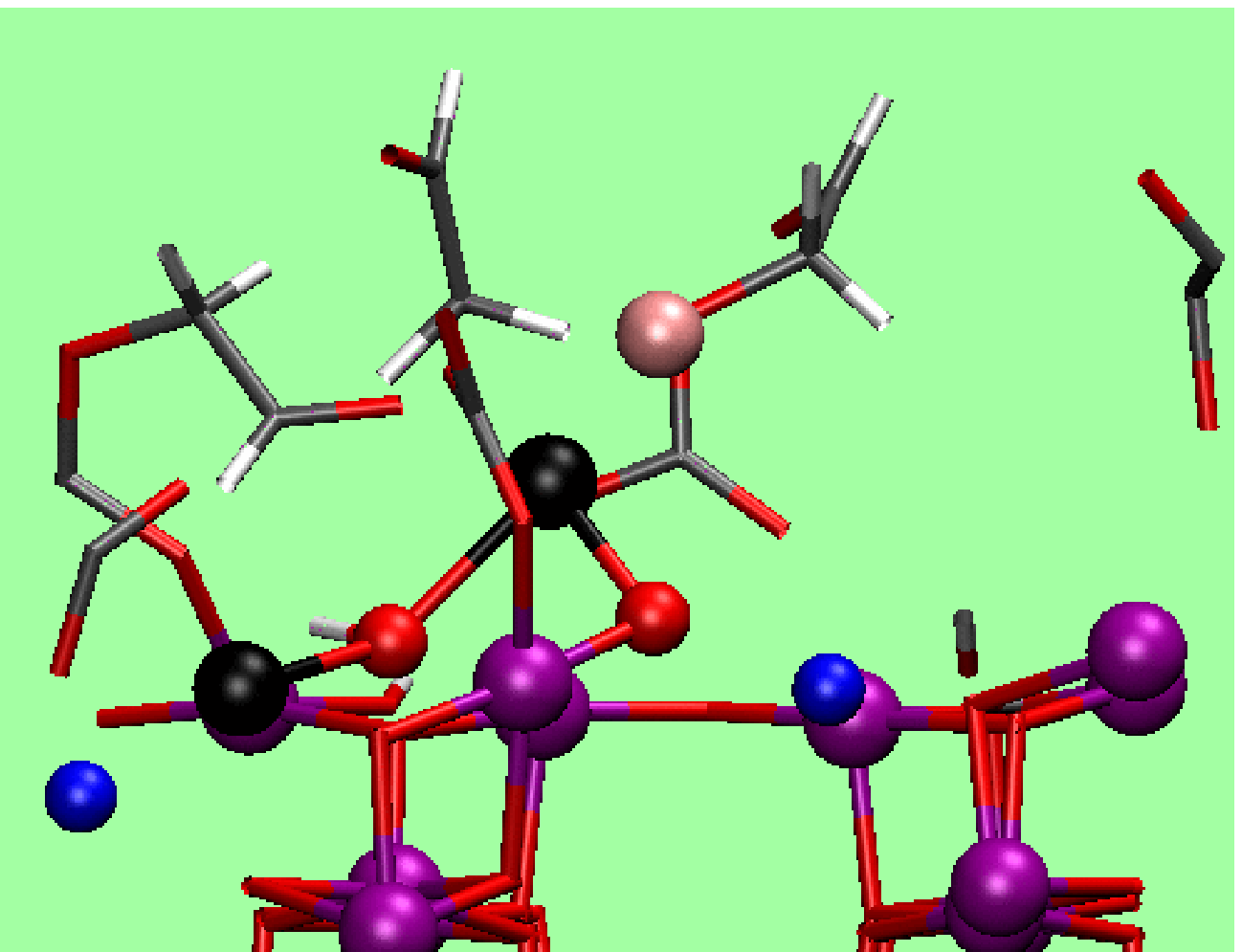}  
                   \epsfxsize=2.00in \epsfbox{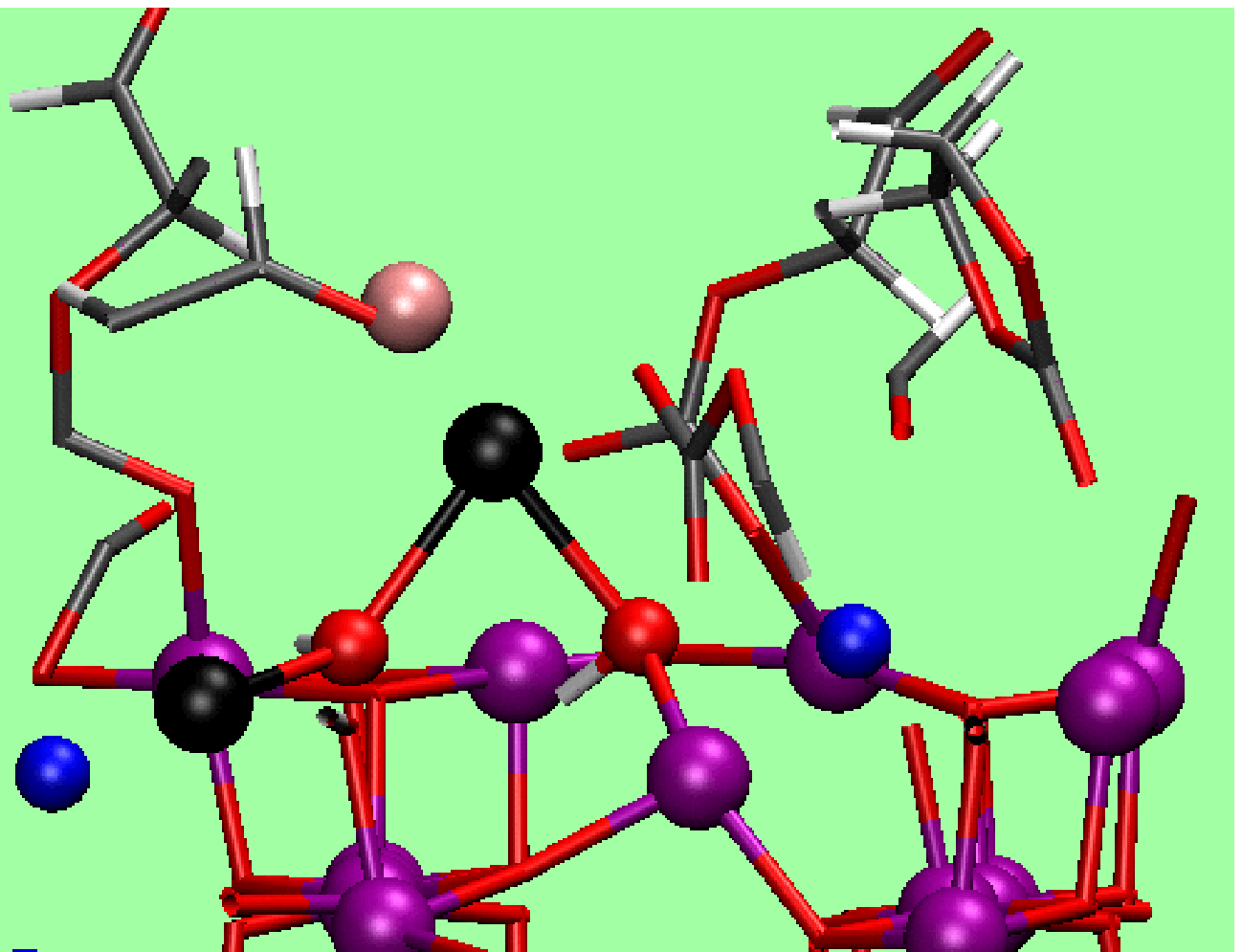} (b)}}
\centerline{\hbox{ (c) \epsfxsize=2.00in \epsfbox{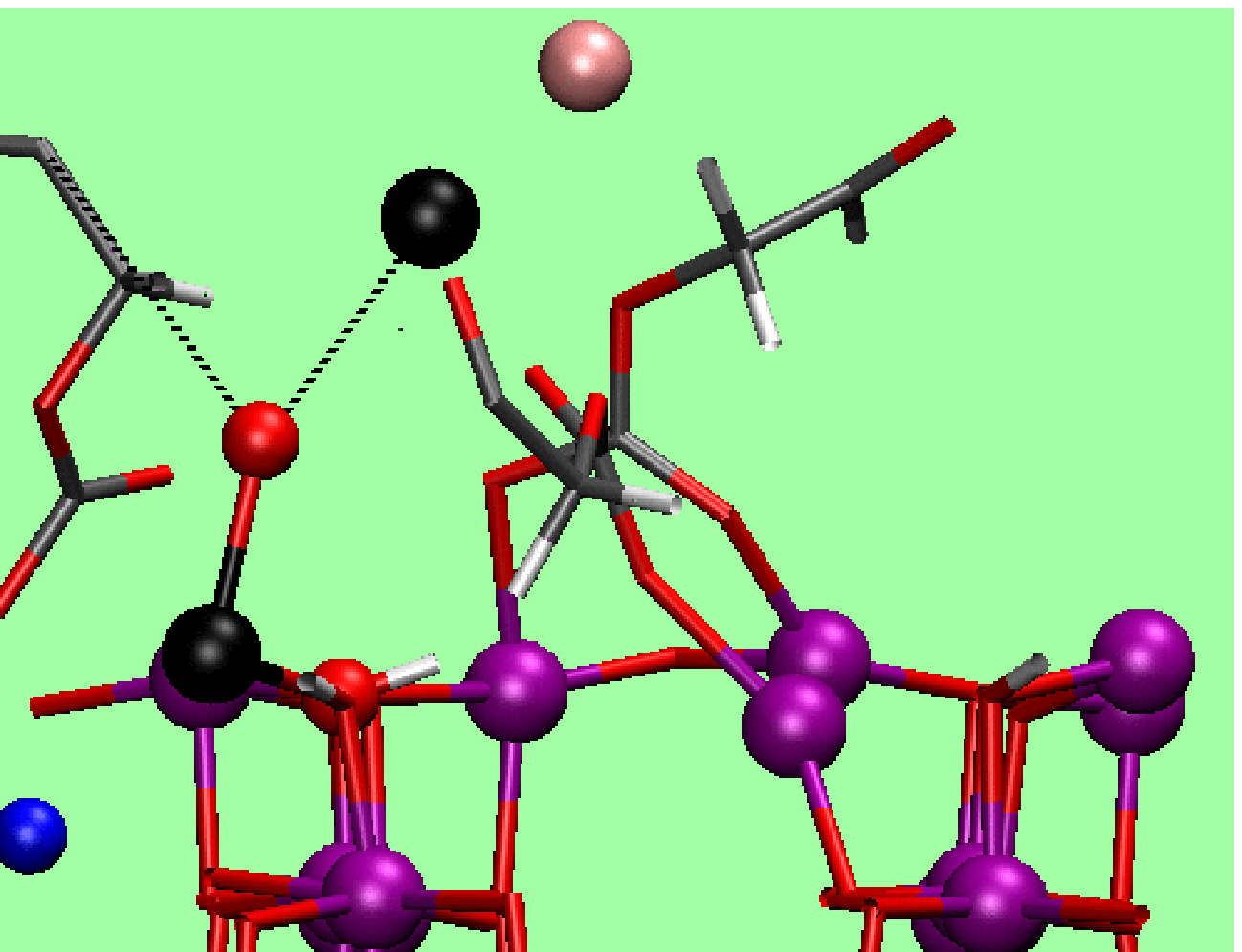}  
                   \epsfxsize=2.00in \epsfbox{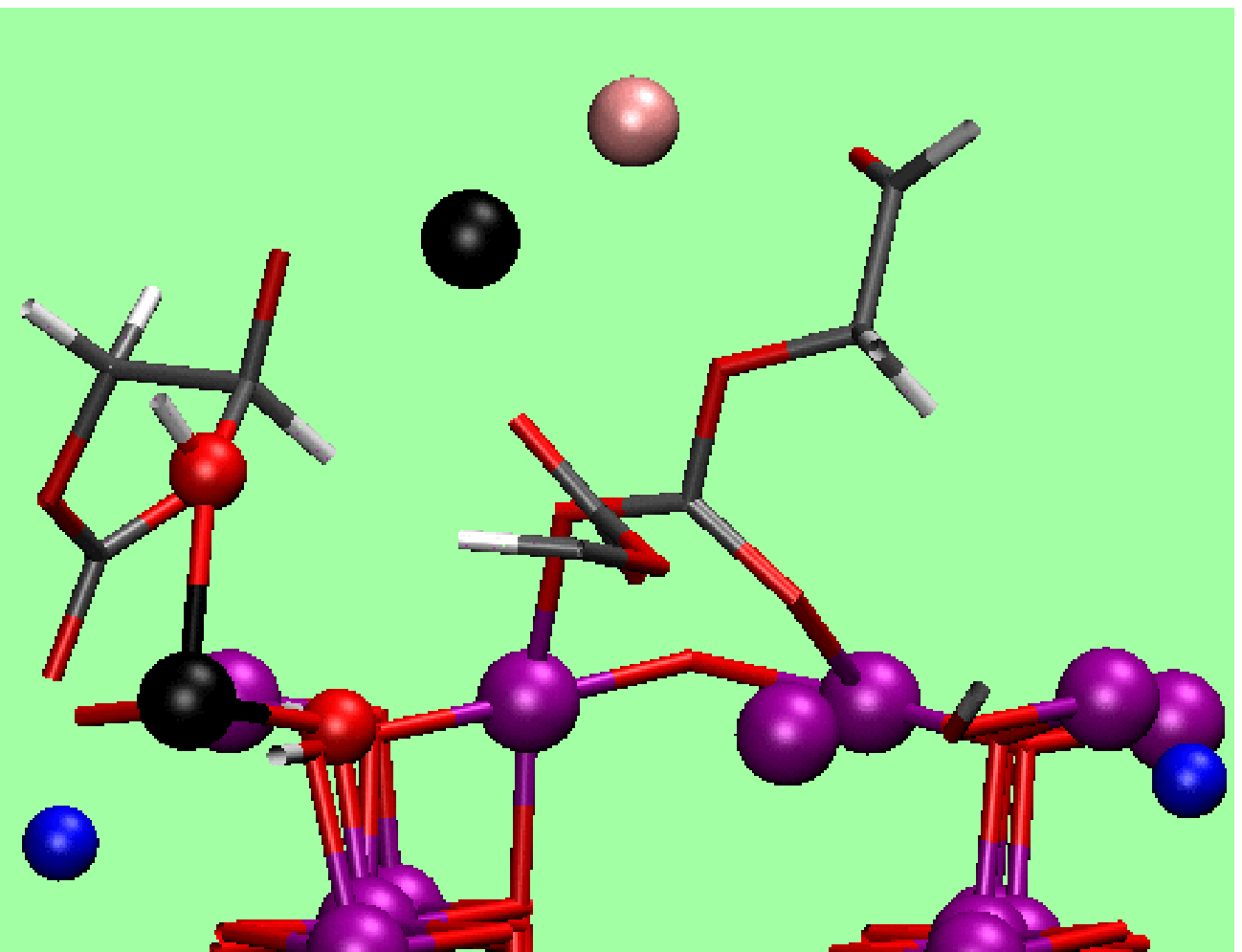} (d)}}
\caption[]
{\label{fig5} \noindent
(a)-(d) Snapshots along the Mn(II) dissolution reaction
coordinate $R'$.  Panel (a) depicts the configuration in Fig.~\ref{fig3}e
rotated by 90~degrees.  In (b), a proton is manually moved to the O$^{2-}$
bonded to the surface Mn(II), and the configuration is re-equilibrated.
The O atoms of the two relevant OH$^-$ groups, only, are depicted as spheres.
The H-atom of one of the these OH$^-$ groups is obscured in panel (c).  The
color scheme is as in Fig.~\ref{fig3}.  In addition, both the tagged
Mn and the surface Mn used to define the new coordinate $R'$ are in black.
}
\end{figure}

We start with the unconstrained configuration shown in Fig.~\ref{fig3}e, and
switch to a different reaction coordinate, $R'=|z_{\rm Mn}-z_{\rm Mn'}|$, where
(Mn') is another Mn on the (001) surface layer (Fig.~\ref{fig5}a).  
In most of the sampling windows in this PMF calculation, the tagged Mn
is in the Mn(II) charge state.  This Mn ion is initially coordinated to a
surface O$^{2-}$ and a surface OH$^-$ in Fig.~\ref{fig5}a, in addition to
a F$^-$ and a decomposed EC fragment.  To accelerate dissolution, we manually
move a H$^+$ from another OH$^-$ group on the same surface to the O$^{2-}$
coordinated to that Mn.  The exiting Mn is now coordinated to two surface
OH$^-$ groups (Fig.~\ref{fig5}b).  This mimics the effect of either proton
migration or further solvent degradation-induced donation of H$^+$ to the
surface.  Electrostatically, it is reasonable that H$^+$ preferentially binds
to an O$^{2-}$ bridging a Mn(II) and a Mn(III) ion rather than an O$^{2-}$
bridging two Mn(III) on the surface.

\begin{figure}
\centerline{\hbox{ \epsfxsize=4.50in \epsfbox{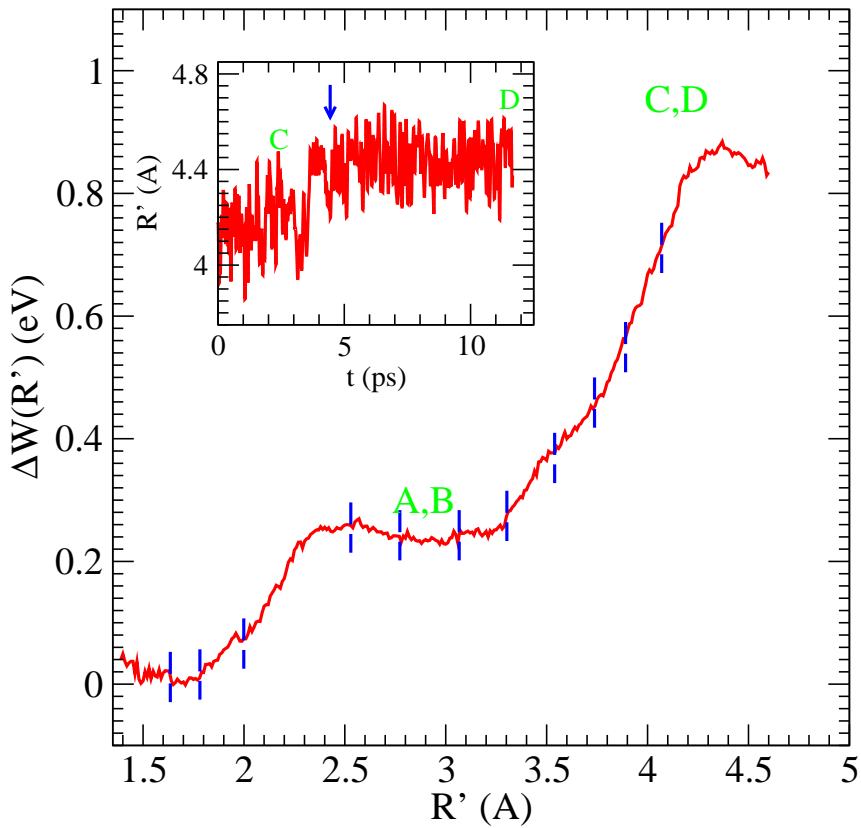} }}
\caption[]
{\label{fig6} \noindent
Potential-of-mean-force for Mn(II) dissociation from the above-surface
LMO site as a function of the new reaction coordinate $R'$.  Inset: Reaction
coordinate ($R'$) as a function of time in the rightmost sampling window
where Mn(II) is released into the liquid electrolyte.  The arrow indicates
the approximate time of OH$^-$ attack on an organic fragment (see text).
A-D refer to panels in Fig.~\ref{fig5}.
}
\end{figure}

Figs.~\ref{fig5}c-d are taken from a trajectory in a sampling window with
an umbrella constraint centered around $R_o=4.4$~\AA.  The average free energy
of this window is $\sim$0.88~eV (Fig.~\ref{fig6}).  These panels depict the
before-and-after snapshots of an unexpected reaction that accompanies Mn(II)
release into the liquid electrolyte.  In Fig.~\ref{fig5}c, taken at the
beginning of the trajectory, one of the two
Mn(II)-OH$^-$ bonds is already broken.  At this stage, AIMD simulations have
led to substantial OH$^-$ re-arrangement.  Comparison with Fig.~\ref{fig5}b
reveals that the OH$^-$ detached from the Mn(II) has moved more than 2~\AA\,
along the surface.  In fact, the $R'< 3$~\AA\, part of (Fig.~\ref{fig6}) is
recomputed by starting from such a OH$^-$ displaced configuration and reducing
the constraint distance $R_o$ progressively in new sampling windows.

The remaining OH$^-$ tethering the Mn(II) to a surface Mn(III) is in the
vicinity of an EC fragment adsorbed to the surface (Fig.~\ref{fig5}c).  3.6~ps
into the trajectory associated with this window, the Mn(II) dissociates from
this OH$^-$.  At 4.9~ps, the Mn(II)-OH distance shrinks back
to 2.9~\AA.  At that point, instead of reforming the Mn(II)-OH$^-$ ionic bond,
the OH$^-$ attacks the -CHO group of the EC fragment nearby to form a
-CH(OH)(O$^-$) motif.  The final configuration is depicted in Fig.~\ref{fig5}d,
taken 11~ps into the trajectory.  The unusual species, like a -COOH group
attacked by a H$^-$, is not expected to be extremely stable.  Indeed the C-OH
covalent bond has a tenuous $\sim$1.55~\AA\, bond length.  Its formation
reflects the strong nucleophilic nature of OH$^-$ weakly solvated by an
aprotic liquid electrolyte.  

Once the OH$^-$ attacks the organic fragment, it is released from both the LMO
surface and the tagged Mn(II) (Fig.~\ref{fig5}d).  The Mn(II)-OH$^-$-Mn(III)
bridge is permanently broken.  The inset to Fig.~\ref{fig6} indeed
shows that, after the OH$^-$ attack, $R'$ fluctuates around 4.4~\AA\, which is
the precisely the umbrella constraint distance $R_o$ in this window.  Therefore
the Mn(II) is undergoing free diffusion, constrained only by the
PMF harmonic potential.  The C-OH bond formation is not reversible
within our AIMD trajectory timescales.  Not even the metadynamics
technique\cite{meta} could have accelerated the C-OH bond formation because
no bond is broken, and the reaction relies on diffusion of the EC fragment.
Thus Fig.~\ref{fig6} does not truly represent a reversible work.  We regard
the predictions of this subsection as semi-quantitative.  However, the
mechanistic steps described herein may be generally applicable to transition
metal ion dissolution in organic solvents.

\subsection{Mn(II)-assisted decomposition of LEDC}

Finally, we explore possible capacity-fade mechanisms caused by Mn(II)
corruption of the anode SEI.  This section originates from our attempt to
compare the binding energy of Mn(I) with the $\beta$-diketonate of
Ref.~\onlinecite{jarry}, and the cohesive energy between Mn(I) and the
EC decomposition fragment used to decorate LMO (001) in this work.  When
taken out of the LMO surface and an $e^-$ is added, the Mn(II)/EC-fragment
complex spontaneously decomposes.  This suggests that similar redox
decompsition reaction may also occur in anode SEI organic components, like EDC.

We create a periodically replicated interface model where two layers of LEDC
are placed on a 4-layer Li$_2$CO$_3$ (001) surface slab.  In view of
Refs.~\onlinecite{shkrob} and~\onlinecite{curtiss}, a Mn(II) is placed 
between these two components; it replaces a Li$^+$ ion on the Li$_2$CO$_3$
surface in contact with LEDC.  An excess $e^-$ is also added, making the
simulation cell charge-neutral.  A high-spin spin-state appropriate to
Mn(I) is imposed, and a combination of AIMD simulations and geometry
optimization calculations are conducted.  Instead of a Mn(I) ion, the
procedure generates a locally stable structure in which Mn(II) persists
and is 5-coordinated.  The excess $e^-$ is localized on the CO$_3$ end 
of an EDC coordinated to the Mn(II), which adopts a characteristic $sp^3$
hybridization.  See the S.I.~for details.

\begin{figure}
\centerline{\hbox{ (a) \epsfxsize=2.00in \epsfbox{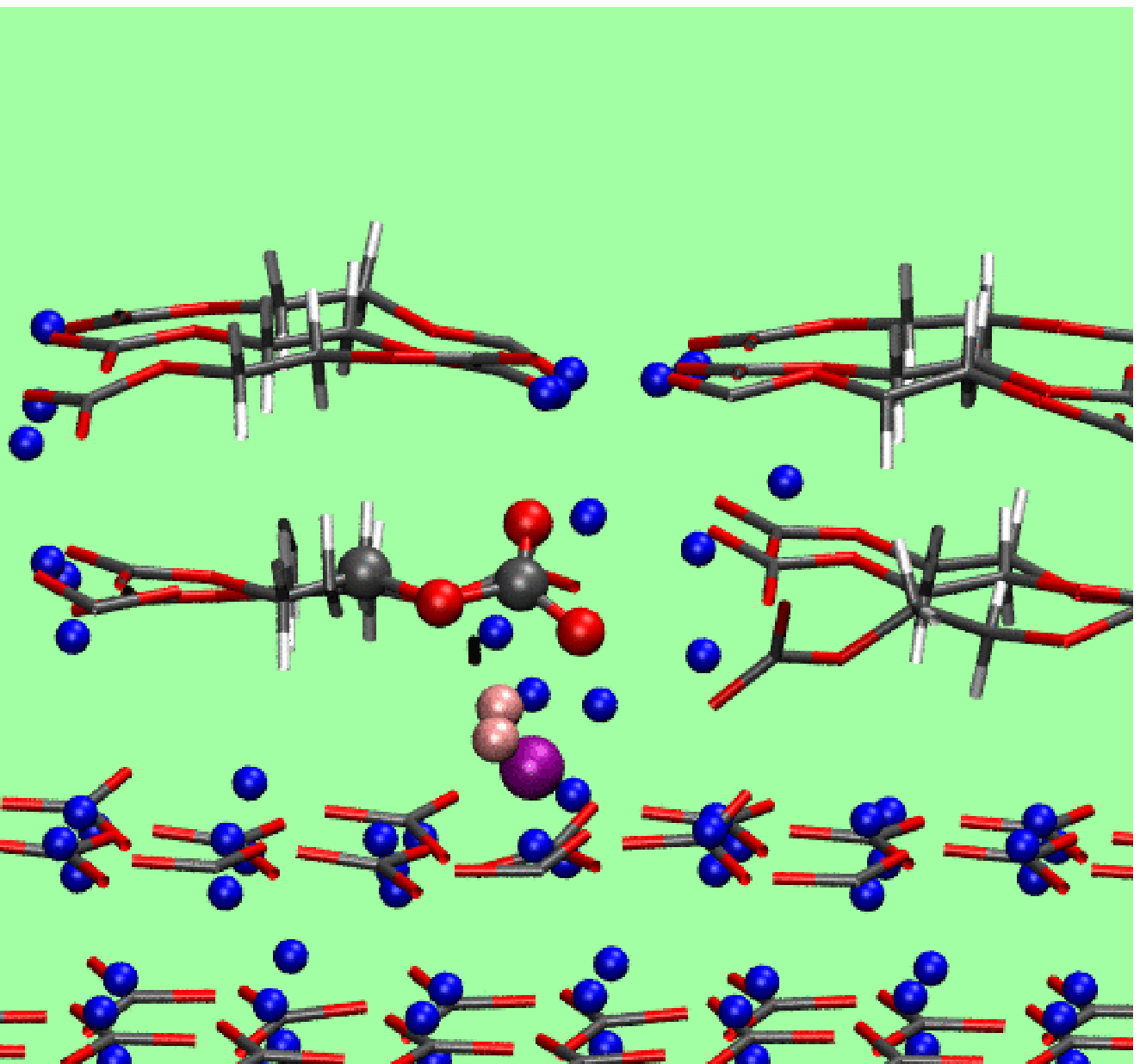}  
                   \epsfxsize=2.00in \epsfbox{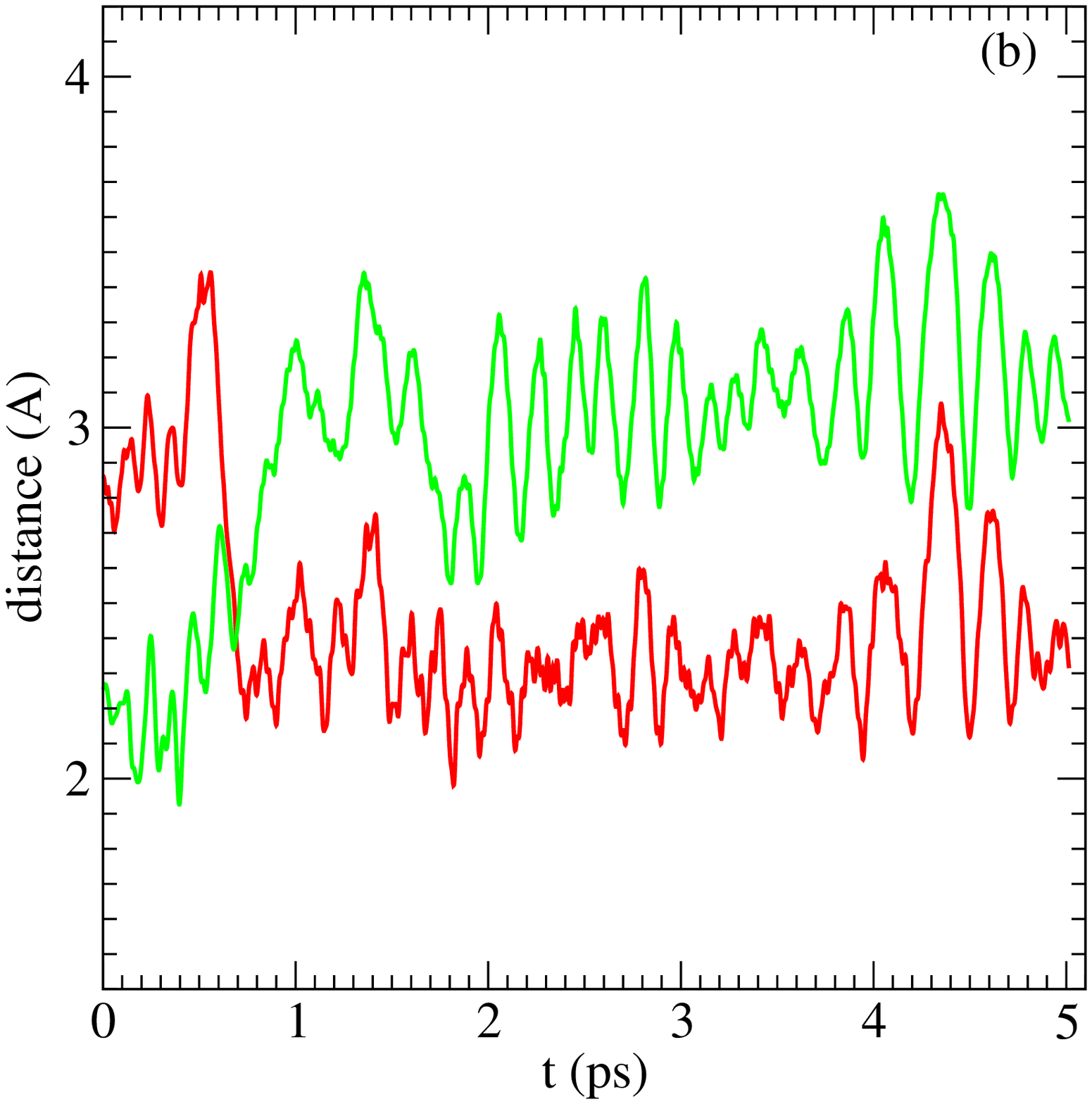} (b)}}
\centerline{\hbox{ (c) \epsfxsize=2.00in \epsfbox{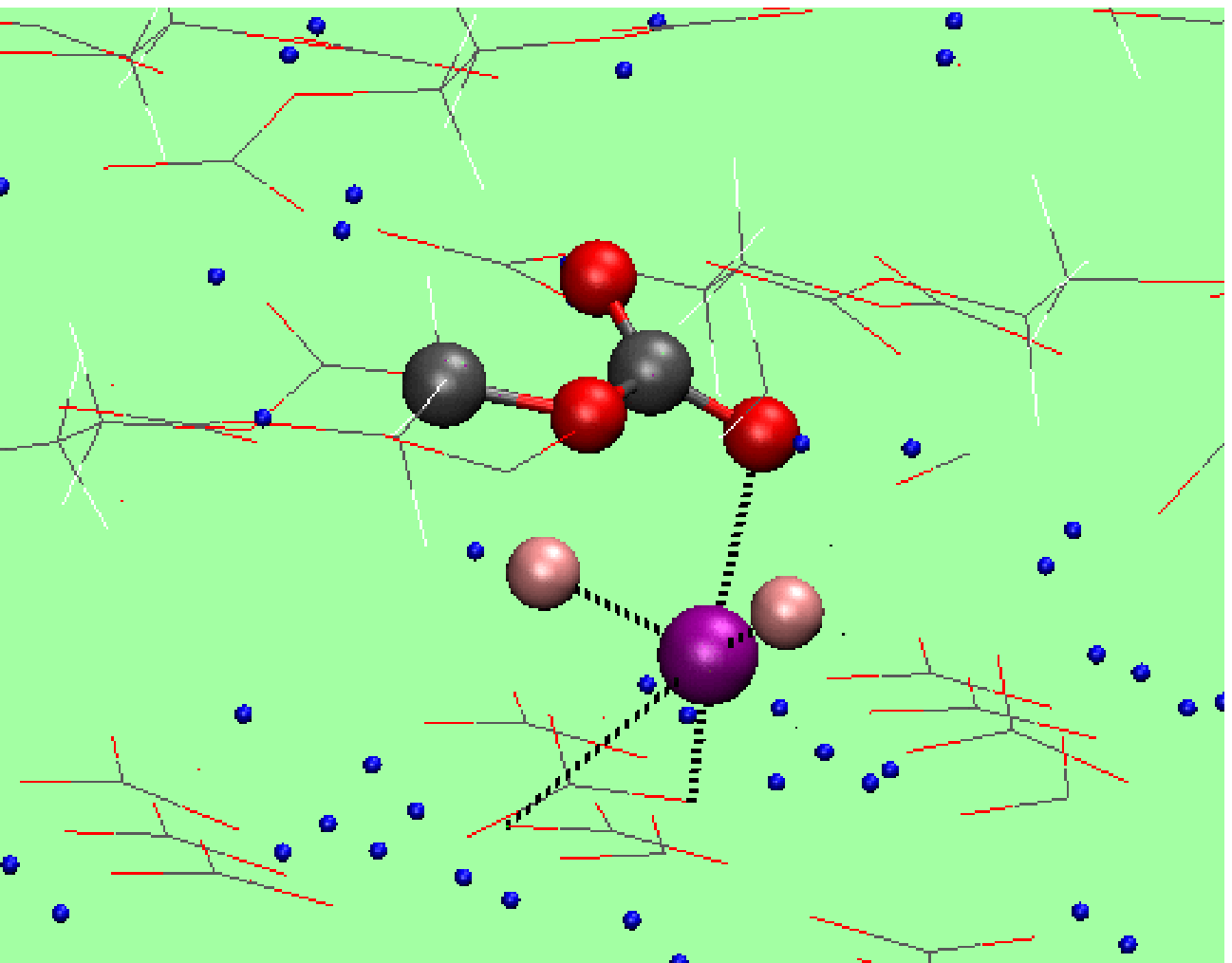}  
                   \epsfxsize=2.00in \epsfbox{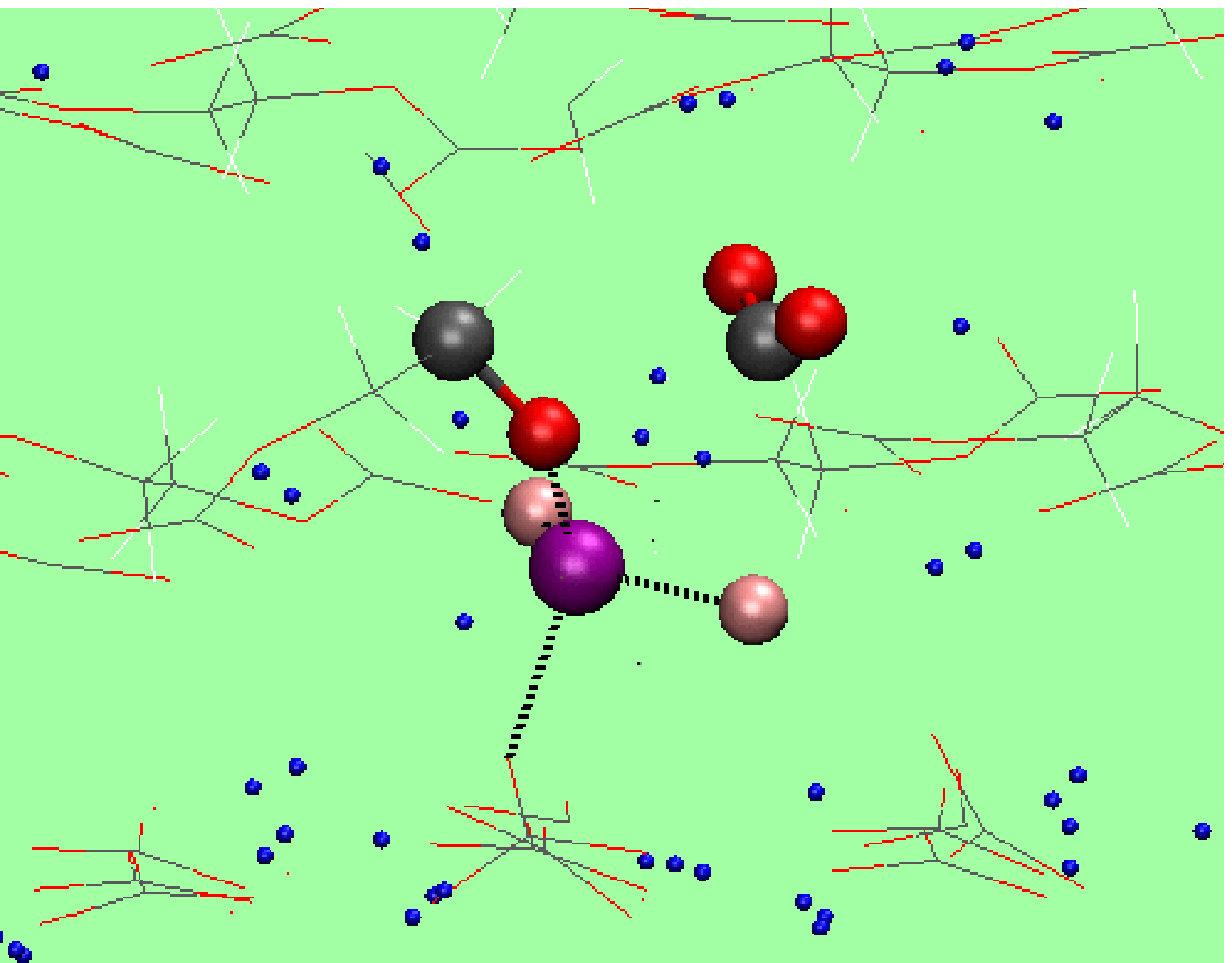} (d)}}
\caption[]
{\label{fig7} \noindent
(a)-(d) Snapshots along an AIMD trajectory of a Mn(II) and an excess $e^-$
at a LEDC/Li$_2$CO$_3$ interface.  Panel (b) is a close-up of (a). (c) and
(d) are taken 0.5~ps and 5.0~ps into the trajectory.  Mn(II), F$^-$, and
the reacting functional group are depicted as spheres while Li$^+$ are dots.
In (d), the CO$^-$ group remains attached to its original parent EDC while
the CO$_2^-$ has been released.
}
\end{figure}

In an attempt to create a 6-coordinated Mn(II), two LiF dimers are inserted
near the Mn(II) ion, and geometry optimization is re-initiated.  Even after
this addition, the Mn(II) remains 5-coordinated, moving 2.55~\AA\, away
from one of the O-atoms initially coordinated to it before
LiF are introduced (Fig.~\ref{fig7}a-b).  The reason may be the large
concentration of negative charges surrounding Mn(II) in an interfacial site
that is not well-stablized by long-range Madelung forces.  The excess electron
is now delocalized over the simulation cell.

Starting from the Fig.~\ref{fig7}b configuration, AIMD simulations are
conducted at T=350~K for 5~ps.  Within 0.5~ps, a CO$_2^-$ radical anion
is released from the R-CO$_3^{-}$ end of a EDC bonded to Mn(II).  Both this
species and the RCHO$^-$ remnant are coordinated to the Mn(II) for a time
(Fig.~\ref{fig7}c), but diffuse away within another 4.5~ps (Fig.~\ref{fig7}d).
The rapid diffusion of negatively charged species from Mn(II) is somewhat
surprising, even considering the elevated temperature used in the simulations.
This may again reflect the large local concentration of negative charge
surrounding the Mn(II).  While the existence of MnF$_2$ motifs in the SEI
has been suggested,\cite{xiao14} F$^-$ is not necessary for CO$_2^-$
release.  The simulation cell obtained prior to adding the two Li$^+$F$^-$
pairs reacts in an analogous way (S.I.).  

Due to the divalent cation induction effect, the Mn(II)(EDC)$_n$ complex
should more readily accept an $e^-$ from the anode at low voltages than the
rest of the SEI components.  Ref.~\onlinecite{curtiss} proposes that
solvent molecules like EC can diffuse through the porous organic SEI
component (e.g., LEDC), coordinate to Mn(II), and be reductively decomposed.
Our calculations suggest an alternate scenario: further reductive decomposition
of organic SEI components like EDC coordinated to Mn(II) (Fig.\ref{fig1}d).
The CO$_2^-$ radical anion released can readily diffuse through the organic
SEI layer without the need of large pores.  It can subsequently attack the
liquid electrolyte outside the SEI layer, causing more capacity fade.
The EDC molecule that loses a CO$_2$ becomes detached from the Mn(II).  
We have not conducted a sufficiently long AIMD trajectory to discover its
ultimate fate, or to see whether other EDC can diffise to and replenish
the Mn(II) coordination shell, causing continuous CO$_2^-$ release as long
as $e^-$ are available.   

From these calculations, our picture of anode SEI-embedded Mn(II) coordination
structure, function, and mode of transport from cathode is more dynamic than
that of Shkrob {\it et al.}\cite{shkrob}  Two-electron reduction
of Mn(II) at the interface has not been examined in this modeling work.
Other researchers have reported possible Mn metal clusters inside the
anode SEI.\cite{nidiffus,xiao14,gowda}  For completeness, the S.I.~shows
that EDC also decomposes on transition {\it metal} surfaces.

\section{Discussions -- Relevance to Computational Design}
\label{discuss}

A key finding of this work is the sheer complexity of Mn migration and
dissolution mechanisms.  Many moving parts are involved.  This section focuses
on two issues that may help design of novel interfaces and protective
strategies which are more resistant to transition metal dissolution.
They are unexpected chemical reactions with the organic fragments, and
the role of OH$^-$.

Regarding chemical reactions, we first stress that the widely-quoted statement,
that organic carbonate solvents are ``stable'' until the cathode potential
exceeds $\sim$4.5~V, is somewhat misleading.  It refers only to oxidative
or electrochemical stability, not intrinsic thermodynamics.  If one use a
solid-state battery component definition of stability,\cite{mo1,mo2} EC, DMC,
and organic fragments derived therefrom, whether they are on the anode or
cathode, are typically thermodynamically unstable.\cite{batt}
Ref.~\onlinecite{leung1} has pointed out that, by itself (without removing
$e^-$),
\begin{equation}
{\rm EC} \rightarrow {\rm CO}_2 + {\rm CH}_3{\rm CHO} , \label{eq1}
\end{equation}
is exothermic by 0.317~eV at T=0~K, not counting zero-point energy which
further favors the reaction.  It is evidently hindered by slow kinetics at
room temperature.  In contrast, all-solid-state battery components are
generally annealed at elevated temperatures, which facilitates the attainment
of thermodynamic equilibrium.  When EC is allowed to react with even fully
lithiated (i.e., discharged) spinel oxide, the reaction is also
exothermic.\cite{leung1,leung2,note}   

Decomposed EC fragments on LMO surfaces create surface oxygen vacancies and
protonates the cathode surface.\cite{leung1,leung2}  These fragments 
themselves are thermodynamically metastable.  In the vicinity of transition
metal ions, excess $e^-$/holes, and/or OH$^-$, these comparatively
high-energy-content fragments may undergo further unexpected reactions unless
the reaction barrier is high.  Therefore one attribute of an ideal cathode
surface is either a {\it kinetic} stability towards organic electrolyte
decomposition altogether, or sacrificial inactivation of electrocatalytic
centers on cathode surfaces in order to prevent organic radical attack
of the cathode oxide.\cite{shkrob16}

Here we focus on the ALD protection strategy instead of new electrolytes
or additives.  Experimental investigations have shown that Mn dissolution is
already much reduced, but is not completely eliminated, by existing ALD
coatings.\cite{xiao,xiao14,dudney13}  It is likely that the ALD
layers will have to crack, develop pores, or otherwise partially decompose
to allow Mn dissolution through them.   More diagnostic studies to
pinpoint ALD film breakdown and SEI formation on cathode ALD films
will be extremely useful.\cite{xiao,jung15,dudney13} Computationally,
an important goal is to elucidate ALD-film breakdown mechanisms.  For each
ALD coating material,\cite{ald1,ald2,ald3,ald4,mauger} one can model
electrolyte decomposition under ultrahigh vacuum conditions, with a
judiciously chosen electron sink in the simulation cell to mimic cathode
charging conditions.  Such calculations should provide insights about the
most susceptible degradation pathways that need to be mitigated.

Regarding restricting OH$^-$ activation of metal dissolution: cathode oxides
or protective coating materials based on oxide tend to form OH$^-$ in the
presence of H$^+$.  Reducing the H$_2$O content in the electrolyte and using 
solvent molecules that do not donate H$^+$ to cathode surfaces are clearly
ideal.  In addition, some material surfaces, like TiO$_2$,\cite{selloni} are
more resistant to reacting with H$_2$O to form OH$^-$ groups than
others.\cite{tio2}  We also propose that metal ions with 3+ and 4+ formal
charges in the protective coating layers will raise OH$^-$ migration barriers.
This is significant because, in this work, OH$^-$ migration is shown to
facilitate metal dissolution on LMO not coated by ALD layers.
This observation appears consistent with anecdotal rankings of successful ALD
oxide coatings in the literature.\cite{ald0,mauger,dudney13}  One exception is
the divalent-metal ion-based magnesium {\it fluoride} coating.  It has proved
promising in passivating LNMO, although capacity fade still occurs in less
than 100 cycles at 45$^o$C.\cite{mgf2} Note that there are competing desirable
attributes for protective coatings, like high Li$^+$ mobility, resistance
towards cracking, and others.\cite{wolverton14}  

\section{Conclusions}
\label{conclusion}

We have applied {\it ab initio} molecular dynamics (AIMD) simulations
to investigate Mn loss from spinel Li$_x$Mn$_2$O$_4$ (001) (``LMO'') surfaces.
We distinguish between Mn migration to above-surface, non-crystallographic
sites to form Mn-decomposed EC fragment complexes, and Mn(II) dissolution 
from these surface sites.  We also explore the consequence of Mn(II) lodged
between the organic and inorganic layers of the solid electrolyte interphase
(SEI) after it has diffused through the separator.

When the exiting Mn is bonded to a F$^-$, which can come from PF$_6^-$
decomposition, we predict the formation of an above-the-surface Mn
intermediate which is in the Mn(II) charge states a majority of the time.
This is in qualitative agreement with the enhancement of Mn(II) content on
the surface of Ni-doped LMO during battery charging.\cite{wlyang}  A
$\Delta G$=1.25$\pm$0.09~eV free energy barrier ($\Delta G^*$) is predicted
for this process.  This $\Delta G^*$, consistent with years of reaction time,
is slightly overestimated, perhaps due to DFT inaccuracies.  Mechanistically,
it is found that concerted solid-state Li$^+$ and liquid electrolyte
motions at the interface facilitate Mn(II) migration to the above the surface.  

The next step -- Mn(II) dissolution from the surface -- must be discussed
semi-quantitatively.  We assume that the Mn which has migrated above the
LMO surface is stabilized by external means and restart AIMD free energy
calculations.  Dissolution is aided by the existence of sufficient H$^+$
so that the exiting Mn(II) only exhibits OH$^-$ bridges to the oxide surface.
OH$^-$ attack on an organic fragment is also observed in our simulations.
Future work will examine whether such an attack is a general phenomenon.
The frequently quoted Hunter mechanism of surface Mn(III) disproporionation
to form Mn(II) is only one early step in a complex process.

Finally, Mn(II) lodged between the organic (LEDC) and inorganic (Li$_2$CO$_3$)
regions of the SEI covering the graphite anode surface can readily decompose EDC
molecules coordinated to it to give CO$_2^-$ radical anions if excess electrons
arrive from the anode.  CO$_2^-$ can diffuse through the organic SEI region,
without requiring large pores to exist, and then attack the liquid electrolyte
outside the SEI, leading to capacity fade.  Our finding that the organic SEI
on the anode surface can undergo chemical reactions dovetails with recent
computational work focused on SEI instability.\cite{batt}  It is also
related to the ``redox shuttle'' route of $e^-$ transport
through the anode SEI.\cite{tang}

In terms of computation, AIMD simulations of the free energy barrier 
associated with Mn loss prove to be challenging.  Concerted solid- and
liquid-state motion usually requires different simulation time scales, and
many moving parts and unexpected chemical reactions can occur while the
transition metal ion moves through the interface.  Given the dearth of
atomic-length-scale experimental interfacial structures as starting points
of simulations, our models should be considered plausible scenarios that can
yield useful insights.

\section*{Acknowledgement}
 
We thank Ilya Shkrob for extensive discussions and semi-empirical calculations
on Mn(II)-catalyzed electrochemical reduction of LEDC in the anode SEI, and
acknowledge Nancy Missert and Ilya Shkrob for their critical reading of the
manuscript.  We also thank Angelique Jarry and Wanli Yang for sharing their
unpublished results and valuable suggestions.

Sandia National Laboratories is a multimission laboratory
managed and operated by Sandia Corporation, a wholly owned subsidiary of
Lockheed Martin Corporation, for the U.S.~Deparment of Energy's National
Nuclear Security Administration under contract DE-AC04-94AL85000.  
This work was supported by Nanostructures for Electrical
Energy Storage (NEES), an Energy Frontier Research Center funded by
the U.S.~Department of Energy, Office of Science, Office of Basic Energy
Sciences under Award Number DESC0001160.

\section*{Supporting Information Available}
Further information are provided on simulations of Mn(II) surface
migration in the absence of F$^-$; EDC decomposition in the absence
of F$^-$; EDC decomposition on the Ni(111) surface; and H$_2$O attack
on the LMO (111) surface.
This information is available free of charge via the Internet
at {\tt http://pubs.acs.org/}.


\begin{references}

\bibitem{thackeray_rev}
Thackeray, M.M.  Manganese Oxides for Lithium Batteries.
{\it Prog. Solid State Chem.} {\bf 1997}, {\it 25}, 1-71.

\bibitem{ram_spinel}
Manthiram, A.; Chemelewski, K.; Lee, E.-S.
A Perspective on the High-Voltage LiMn$_{1.5}$Ni$_{0.5}$O$_4$ Spinel Cathode
for Lithium-Ion Batteries.  
{\it Energy Environ.~Sci.} {\bf 2014}, {\it 7}, 1339-1349.

\bibitem{jow}
Jow, T.R.; Allen, J.L.; Borodin, O.; Delp, S.A.; Allen, J.L.
Challenges in Develoiping High Energy Density Li-Ion Batteries with
High Voltage Cathodes.  {\it TMS 2014 Supp.~Proc.} {\bf 2014}, 853-857.

\bibitem{mnsei1}
Blyr, A.; Sigala, C.; Amatucci, G.; Guyomard, D.; Chabre, Y.; Tarascon, J.-M.
Self-discharge of LiMn$_2$O$_4$/C Li-Ion Cells in their Discharged State --
Understanding by Means of Three-electrode Measurements.
{\it J. Electrochem. Soc.} {\bf 1998}, {\it 145}, 194-209.

\bibitem{mnsei2}
Tsunekawa, H.; Tanimoto, S.; Marubayahsi, R.; Fujita, M.; Kifune, K.; Sano, M.
Capacity Fading of Graphite Electrodes Due to the Deposition of
Manganese Ions on Them in Li-Ion Batteries.
{\it J. Electrochem. Soc.} {\bf 2002}, {\it 149}, A1326-1331.

\bibitem{mnsei3}
Amine, K.; Liu, J.; Kang, S.; Belharouak, I.; Hyang, Y.; Vissers, D.;
Henriksen, G.  Improved Lithium Manganese Oxide Spinel/Graphite Li-Ion
Cells for High Power Applications.
{\it J. Power Sources}, {\bf 2004}, {\it 129}, 14-19.

\bibitem{mnsei4}
Amine, K; Liu, J.; Belharouak, I.; Kang, S.-H.; Bloom. I.; Vissers, D.; 
Henriksen, G. Advanced Cathode Materials for High-Power Applications.
{\it J. Power Sources}, {\bf 2005}, {\it 146}, 111-115.

\bibitem{mnsei5}
Yang, L.; Takahashi, M.; Wang, B.F.   A Study of Capacity Fading of Lithium-Ion
Battery with Manganese Spinel Positive Electrode during Cycling.
{\it Electrochim. Acta} {\bf 2006}, {\it 51}, 3228-3234.

\bibitem{mnsei6}
Delacourt, C.; Kwong, A.; Liu, X.; Qiao, R.; Yang, W.L.; Lu, P.; Harris, S.J.;
Srinivasan, V.  Effect of Manganese Contamination on the
Solid-Electrolyte-Interphase Properties in Li-Ion Batteries.
{\it J.~Electrochem.~Soc.} {\bf 2013}, {\it 160}, A1099-1107.

\bibitem{ram06}
Choi, W.; Manthiram, A. Comparison of Metal Ion Dissolutions from Lithium
Ion Battery Cathodes.  {\it J.~Electrochem.~Soc.} {\bf 2006}, {\it 153}, 
A1760-A1764.

\bibitem{spinelht1}
Matsuo, Y.; Kostecki, R.; McLarnon, F.  
Surface Layer Formation on Thin-Film LiMn$_2$O$_4$ Electrodes at Elevated
Temperatures.  {\it J. Electrochem. Soc} {\bf 2001}, {\it 148}, A687.

\bibitem{spinelht2}
Du Pasquier, A.; Blyr, A.; Courjal, P.; Larcher, D.; Amatucci, G.;
G\'{e}rand, B.; Tarason, J.-M.  
Mechanism for Limited 55$^o$C Storage Performance of Li$_{1.0}$Mn$_{1.95}$O$_4$
Electrodes.  {\it J. Electrochem. Soc} {\bf 1999}, {\it 146}, 428-436.

\bibitem{spinelht3}
Du Pasquier, A.; Blyr, A.; Cressent, A.; Lenain, C.; Amatucci, G.; Tarason,
J.-M.  An Update on the High Temperature Ageing Mechanism in
LiMn$_2$O$_4$-based Li-ion Cells.
{\it J.~Power Sources} {\bf 1999}, {\it 81-82}, 54-59.

\bibitem{tarascon91}
Tarascon, J.M.; Wang, E.; Shokoohi, F.K.; McKinnon, W.R.; Colson, S.
The Spinel Phase of LiMn$_2$O$_4$ as a Cathode in Secondary Lithium Cells.
{\it J.~Electrochem.~Soc.} {\bf 1991}, {\it 138}, 2859-2864.

\bibitem{xiao}
Xiao, X.C.; Ahn, D.; Liu, Z.; Kim, J.-H.; Lu, P.
Atomic Layer Coating to Mitigate Capacity Fading Associated with
Manganese Dissolution in Lithium Ion Batteries.
{\it Electrochem.~Commun.} {\bf 2013}, {\it 32}, 31-34.

\bibitem{jung15}
Kim, J.W.; Kim, D.H.; Oh, D.Y.;  Lee, H.; Kim, J.H.; Lee, J.H.; Jung, Y.S.
Surface Chemistry of LiNi$_{0.5}$Mn$_{1.5}$O$_4$ Particles Coated by
Al$_2$O$_3$ Using Atomic Layer Deposition for Lithium-Ion Batteries.
{\it J.~Power~Sources}, {\bf 2015}, {\it 274}, 1254-1262.

\bibitem{ald0}
Jung, Y.S.; Cavanagh, A.S.; Dillon, A.C.; Groner, M.D.;
George, S.M.; Lee, S.-H. 
Enhanced Stability of LiCoO$_2$ Cathodes in Lithium-Ion
Batteries Using Surface Modification by Atomic Layer Deposition.
{\it J.~Electrochem.~Soc.} {\bf 2010}, {\it 157},  A75-A81.

\bibitem{nidiffus}
Joshi,~T.; Eom, K.; Yushin, G.; Fuller, T.F.
Effects of Dissolved Transition Metals on the Electrochemical Performance
and SEI Growth in Lithium-Ion Batteries.  {\it J.~Electrochem.~Soc.}
{\bf 2014}, {\it 161}, A1915-A1921.

\bibitem{hunter}
Hunter, J.C. 
Preparation of a New Crystal Form of Manganese Dioxide:
$\lambda$-MnO$_2$. {\it J. Solid State Chem.} {\bf 1981}, {\it 39}, 142-147.

\bibitem{uhv}
Ouyang, C.Y.; Sljivancanin, Z.; Baldereschi, A.  Oxidation States of Mn Atoms
at Clean and Al$_2$O$_3$-Covered LiMn$_2$O$_4$ (001) Surfaces.  
{\it Phys. Rev. B} {\bf 2009}, {\it 79}, 235410.

\bibitem{leung1}
Leung,~K.  First-Principles Modeling of the Initial Stages of Organic
Solvent Decomposition on Li$_x$Mn$_2$O$_4$ Surfaces.
{\it J. Phys. Chem. C} {\bf 2012}, {\it 116}, 9852-9861.

\bibitem{kanno1}
Hirayama, M.; Ido, H.; Kim, K.S.; Cho, W.; Tamura, K.; Mizuki, J.; Kanno, R.
Dynamic Structural Changes at LiMn$_2$O$_4$/Electrolyte Interface during
Lithium Battery Reaction.
{\it J. Am. Chem. Soc.} {\bf 2010}, {\it 132}, 15268-15276.

\bibitem{kanno2}
Hirayama, M.; Sonoyama, N.; Ito, M.; Minoura, M.; Mori, D.; Yamada, A.; 
Tamura, K.; Mizuki, J.; Kanno, R.  
Characterization of Electrode/Electrolyte Interface with X-ray Reflectometry
and Epitaxial-Film LiMn$_2$O$_4$ Electrode.
{\it J. Electrochem.~Soc.} {\bf 2007}, {\it 154}, A1065-1072.

\bibitem{song}
Song, J.-W.; Ngyuen, C.C.; Choi, H.; Lee, K.-H.; Han, K.-H.; Kim, Y.-J.;
Choy, S.; Song, S.W. 
Impacts of Surface Mn Valence on Cycling Performance and Surface Chemistry
of Li- and Al-Substituted Spinel Battery Cathodes.
{\it J.~Electrochem.~Soc.} {\bf 2011}, {\it 158}, A458-A464.

\bibitem{oh}
Yoon, T.; Kim, D.; Park, K.H.; Park, H.; Jurng, S.; Jang, J.; Ryu, J.H.; 
Kim, J.J.; Oh, S.M.
Composition Change of Surface Film Deposited on LiNi$_{0.5}$Mn$_{1.5}$O$_4$
Positive Electrode. {\it J.~Electrochem.~Soc.} {\bf 2014}, {\it 161}, A519-A523.

\bibitem{novak}
Simmen, F.; Hintennach, A.; Horisberger, M.; Lippert, T.; Nov\'{a}l, P.;
Schneider, C.W.; Wokaun, A.  Aspects of the Surface Layer Formation on
Li$_{1+x}$Mn$_2$O$_{4-\delta}$ during Electrochemical Cycling.
{\it J. Electrochem. Soc.} {\bf 2010}, {\it 157}, A1026-A1029.

\bibitem{edstrom}
Edstr\"{o}m, K.; Gustafsson, T.; Thomas, J.O.
The Cathode-Electrolyte Interface in the Li-ion Battery.
{\it Electrochem.~Acta} {\bf 2004}, {\it 50}, 397-403.

\bibitem{aurbach01}
Moshkovich, M.; Cojocaru, M.; Gottlieb, H.E.; Aurbach, D.
The Study of the Anodic Stability of Alkyl Carbonate Solutions by
in situ FTIR Spectroscopy, EQCM, NMR, and MS.
{\it J. Electroanal. Chem.} {\bf 2001}, {\it 497}, 84-96.

\bibitem{aurbach99}
Aurbach, D.; Markovsky, B.; Levi, M.D.; Levi, E.; Schechter, A.; Moshkovich,
M.; Cohen, Y.  New Insights into the Interactions between Electrode
Materials and Electrolyte Solutions for Advanced Nonaqueous Batteries.
{\it J. Power Sources} {\bf 1999}, {\it 81}, 95-111.

\bibitem{doeff}
Lin, F.; Markus, I.M.; Nordlund, D.; Weng, T.-C.; Asta, M.D.; Xin, H.L.; 
Doeff, M.M.  Surface Reconstruction and Chemical Evolution of Stoichiometric
Layered Cathode Materials for Lithium-Ion Batteries.  {\it Nat.~Commun.}
{\bf 2014}, {\it 5}, 1-9, and references therein.

\bibitem{wlyang}
Qiao, R.; Wang, Y.; Olalde-Velasco, P.; Li, H.; Hu, Y.-S.; Yang, W.
Direct Evidence of Gradient Mn(II) Evolution at Charged States in
LiNi$_{0.5}$Mn$_{1.5}$O$_4.$  {\it J.~Power Sources}, {\bf 2015}, {\it 273},
1120-1126.

\bibitem{jarry}
Jarry, A.; Gottis, S.; Yu, Y.-S.; Roque-Rosell, J.; Kim, C.; Cabana,
J.; Kerr, J.; Kostecki, R.  The Formation Mechanism of Fluorescent
Metal Complexes at the Li$_x$Ni$_{0.5}$Mn$_{1.5}$O$_{4-\delta}$/Carbonate
Ester Electrolyte Interface.  {\it J.~Am.~Chem.~Soc.} {\bf 2015}, {\it 137},
3533-3539.

\bibitem{leung2}
Kumar, M.; Leung, K.; Siegel, D.J.
Crystal Surface and State of Charge Dependencies of 
Electrolyte Decomposition on LiMn$_2$O$_4$ Cathode.
{\it J.~Electrochem.~Soc.} {\bf 2014}, {\it 161}, E3059-E3065.

\bibitem{borodin2015}
Borodin, O.; Olguin, M.; Spear, C.E.; Leiter, K.W.; Knap, J.
Towards High Throughput Screening of Electrochemical Stability of 
Battery Electrolytes.  {\it Nanotechnology} {\bf 2015}, {\it 26}, 354003.

\bibitem{benedek_dry}
Benedek, B.; Thackeray, M.M. 
Simulation of the Surface Structure of Lithium Manganese Oxide Spinel.
{\it Phys. Rev. B} {\bf 2011}, {\it 83}, 195439.

\bibitem{geochem}
Zhang, C.; Liu, X.; Lu, X.; Meijer, E.J.; Wang, K.; He, M.; Wang, R.
Cadium (II) Complexes Adsorbed on Clay Edge Surfaces: Insight
from First Principles Molecular Dynamics Simulation.
{\it Clays and Clay Minerals}, {\tt http://dx.doi.org/10.1346/CCMN.2016.0640402}

\bibitem{chandler}
Chandler~D.  {\it Introduction to Modern Statistical Mechanics};
Oxford, New York, 1997, Ch.~6.

\bibitem{meta}
Laio, A.; Parrinello, M.  Escaping Free Energy Minima.
{\it Proc. Natl. Acad. Sci. USA} {\bf 2002}, {\it 99}, 12562-12566.

\bibitem{benedek}
Benedek, R.; Thackeray, M.M.; Low, J.; Bucko, T.  
Simulation of Aqueous Dissolution of Lithium Manganate Spinel from First 
Principles.  {\it J. Phys. Chem. C} {\bf 2012}, {\it 116}, 4050-4059.

\bibitem{persson}
Karim, A.; Fosse, S.; Persson, K.A.
Surface Structure and Equilibrium Particle Shape of the
LiMn$_2$O$_4$ Spinel from First-Principles Calculations.
{\it Phys.~Rev.~B} {\bf 2013}, {\it 87}, 075322.

\bibitem{liverpool}
Scivetti, I.; Teobaldi, G.
(Sub)surface-Promoted Disproportionation and Absolute Band Alignment
in High-Power LiMn$_2$O$_4$ Cathodes.
{\it J.~Phys.~Chem.~C} {\bf 2015}, {\it 119}, 21358-21368.

\bibitem{greeley}
Warburton, R.E.; Iddir, H.; Curtiss, L.A.; Greeley, J.
Thermodynamic Stability of Low- and High-Index Spinel LiMn$_2$O$_4$
Surface Terminations. {\it Appl.~Mater.~Interfaces} {\bf 2016}, 
{\it 8}, 11108-11121.

\bibitem{wolverton}
Kim, S.; Aykol, M.; Wolverton, C.
Surface Phase Diagram and Stability of (001) and (111) LiMn$_2$O$_4$
Spinel Oxides.  {\it Phys.~Rev.~B} {\bf 2015}, {\it 92}, 115411.

\bibitem{co3}
Huang, M.-R.; Lin, C.-W.; Lu, H.-Y. 
Crystallographic Facetting in Solid-State Reacted LiMn$_2$O$_4$ Spinel Powder.
{\it Appl. Surf. Sci.} {\bf 2001}, {\it 177}, 103-113.

\bibitem{eriksson}
However, lithium carbonate is not reported in other surface-sensitive studies.
See, e.g., Eriksson, T.; Andersson, A.M.; Bishop, A.G.; Gejke, C.; Gustafsson,
T.; Thomas, J.O.  Surface Analysis of LiMn$_2$O$_4$ Electrodes in
Carbonate-Based Electrolytes.  {\it J. Electrochem. Soc.} {\bf 2002},
{\it 149}, A69-78.

\bibitem{shkrob}
Shkrob, I.A.; Kropf, A.J.; Marin, T.W.; Li, Y.; Poleuktov, O.G.;
Niklas, J.; Abraham, D.P.  Manganese in Graphite Anode and
Capacity Fade in Li Ion Batteries.
{\it J.~Phys.~Chem.~C} {\bf 2014}, {\it 118}, 24335-242348.

\bibitem{curtiss}
Vissers, D.R.; Chen, Z.; Shao, Y.; Engelhard, M.;
Das, U.; Redfern, P.; Curtiss, L.A.; Pan, B.; Liu, J.; Amine, K.
Role of Manganese Deposition on Graphite in the Capacity
Fade of Lithium Ion Batteries.
{\it Appl.~Mater.~Interfaces} {\bf 2016}, {\it 8}, 14244-14251.

\bibitem{graindiffus}
Mn(II) diffusion may however be enhanced in grain boundaries between SEI
components.\cite{yue16}  This is a complex to deal with computationally.

\bibitem{yue16}
Zhang, Q.; Pan, J.; Lu, P.; Liu, Z.; Verbrugge, M.W.; Sheldon, B.W.;
Cheng, Y.-T.; Qi, Y.; Xiao, X.
Synergetic Effects of Inorganic Components in Solid Electrolyte Interphase
on High Cycle Efficiency of Lithium Ion Batteries.
{\it Nano Lett.} {\bf 2016}, {\it 16}, 2011-2016.

\bibitem{ald1}
Liu, J.; Sun, X.
Elegant Design of Electrode and Electrode/Electrolyte Interface in 
Lithium Ion Batteries by Atomic Layer Deposition.
{\it Nanotech.} {\bf 2015}, {\it 26}, 024001.

\bibitem{ald2}
Meng, X.; Yang, X.-Q.; Sun, X.  
Emerging Applications of Atomic Layer Deposition for Lithium-Ion
Battery Studies.  {\it Adv. Mater.} {\bf 2014}, {\it 24}, 3589-3615.

\bibitem{ald3}
Wang, X.; Yushin, G.
Chemical Vapor Deposition and Atomic Layer Deposition for Advanced Lithium
Ion Batteries and Supercapacitors.
{\it Energy Envir.~Sci.} {\bf 2015}, {\it 8}, 1889-1904.

\bibitem{ald4}
Mao, F.; Guo, W.; Ma, J.
Research Progress in Design Strategies, Synthesis, and Performance
of LiMn$_2$O$_4$-based Cathodes.
{\it RSC Advances}, {\bf 2015}, {\it 5}, 105248-105258.

\bibitem{mauger}
Mauger, A.; Julien, C.
Surface Modifications of Electrode Materials for Lithium-Ion Batteries:
Status and Trends. {\bf 2014}, {\it Ionics}, {\it 20}, 751-787.

\bibitem{xiao14}
Xiao, X.C.; Liu, Z.; Baggetto, L.; Veith, G.M.; More, K.L.; Unocic, R.R.
Unraveling Manganese Dissolution/Deposition Mechanisms on the Negative
Electrode Lithium Ion Batteries.
{\it Phys.~Chem.~Chem.~Phys.}, {\bf 2014}, {\it 16}, 10398-10402.

\bibitem{borodin_h+}
Another possible role of PF$_6^-$ is to facilitate organic solvent oxidation,
yielding HF as co-product.  See Xing, L.; Borodin, O.; Smith, G.; Li, W.  
Density Functional Theory Study of the Role of Anions on the Oxidative
Decomposition Reaction of Propylene Carbonate.
{\it J. Phys. Chem.} {\bf 2011}, {\it 115}, 13896.

\bibitem{saulnier}
Saulnier, M.; Auclair, A.; Liang, G.; Schougaard, S.B.
Manganese Dissolution in Lithium-Ion Positive Electrode Materials.
{\it Solid State Ionic}, {\bf 2016}, {\it 294}, 1-5.

\bibitem{sano}
Yamane, H.; Inoue, T.; Fujita, M.; Sano, M.
A Causal Study of the Capacity Fading of Li$_{1.01}$Mn$_{1.99}$O$_4$
Cathode at 80$^o$C, and the Suppressing Substances of Its Fading.
{\it J.~Power~Sources} {\bf 2001}, {\it 99}, 60-65.

\bibitem{oh97}
Jang, D.H.; Oh, S.M. Electrolyte Effects on Spinel Dissolution and Cathodic
Capacity Losses in 4~V Li/Li$_x$Mn$_2$O$_4$ Rechargeable Cells.
{\it J. Electrochem. Soc.} {\bf 1997}, {\it 144}, 3342-3348.

\bibitem{oh12}
Kim, D.; Park, S.; Chae, O.B.; Ryu, J.H.; Kim, Y.-U.; Yin, R.-Z.; Oh, S.M.
Re-Deposition of Manganese Species on Spinel LiMn$_2$O$_4$ Electrode
after Mn Dissolution.
{\it J.~Electrochem.~Soc.} {\bf 2012}, {\it 159}. A193-A197.

\bibitem{borodin}
Borodin, O.; Bedrov, D. 
Interfacial Structure and Dynamics of the Lithium Alkyl Dicarbonate SEI 
Components in Contact with the Lithium Battery Electrolyte.
{\it J.~Phys.~Chem.~C} {\bf 2014}, {\it 114}, 18362-18371.

\bibitem{ochida}
Ochida, M.; Domi, Y.; Doi, Takayuki, Tsubouchi, S.; Nakagawa, H.; Yamanaka, T.;
Abe, T.; Ogumi, Z.
Influence of Manganese Dissolution on the Degradation of Surface Films on
Edge Plane Graphite Negative-Electrodes in Lithium-Ion Batteries.
{\it J.~Electrochem.~Soc.} {\bf 2012}, {\it 159}, A961-A966.

\bibitem{gowda}
Gowda, S.R.; Gallagher, K.G.; Croy, J.R.; Bettge, M.; Thackerary, M.M.;
Balasubramnian, M.  Oxidation State of Cross-over Manganese Species on the
Graphite Electrode of Lithium-Ion Cells.
{\it Phys.~Chem.~Chem.~Phys.}, {\bf 2014}, {\it 16}, 6989-6902.

\bibitem{batt}
Leung, K.; Soto, F.; Hankins, K.; Balbuena, P.; Harrison, K.L.
Stability of Solid Electrolyte Interphase Components on Lithium
Metal and Reactive Anode Material Surfaces.
{\it J.~Phys.~Chem.~C} {\bf 2016}, {\it 120}, 6302-6313.

\bibitem{perla}
Soto, F.; Ma, Y.; Martinez de la Hoz, J.; Seminario, J.; Balbuena, P.B.
Formation and Growth Mechanisms of Solid-Electrolyte Interphase Layers in
Rechargeable Batteries.  {\it Chem. Mater.} {\bf  2015}, {\it 27}, 7990-8000.

\bibitem{interface}
Leung,~K.  Electronic Structure Modeling of Electrochemical Reactions at 
Electrode/Electrolyte Interfaces in Lithium Ion Batteries.
{\it J.~Phys.~Chem.~C} {\bf 2013}, {\it 117}, 1539-1547.

\bibitem{leung3}
Leung,~K.; Leenheer, A.
How Voltage Drops are Manifested by Lithium Ion Configurations at Interfaces
and in Thin Films on Battery Electrodes.
{\it J.~Phys.~Chem.~C} {\bf 2015}, {\it 119}, 10234-10246.

\bibitem{vasp1}
Kresse, G.; Furthm\"{u}ller,~J. 
Efficient Iterative Schemes for Ab Initio Total-Energy Calculations
Using a Plane-wave Basis Set.  {\it Phys.~Rev.~B} {\bf 1996}, {\it 54},
11169-11186. 

\bibitem{vasp1a}
Kresse, G.; Furthm\"{u}ller,~J. 
Efficiency of {\it Ab-initio} Total Energy Calculations for Metals and
Semiconductors using a Plane-Wave Basis Set.
{\it Comput.~Mater.~Sci.} {\bf 1996}, {\it 6}, 15-50.

\bibitem{vasp2}
From Ultrasoft Pseudopotentials to the Projector Augmented-Wave Method.
Kresse~G.; Joubert,~D. {\it Phys.~Rev.~B} {\bf 1999}, {\it 59}, 1758-1775.

\bibitem{vasp3}
Paier,~J.; Marsman,~M.; Kresse,~G. 
Why Does the B3LYP Hybrid Functional Fail for Metals?
{\it J. Chem. Phys.} {\bf 2007}, {\it 127}, 024103.

\bibitem{pbe}
Perdew,~J.P., Burke,~K.;  Ernzerhof,~.M.
Generalized Gradient Approximation Made Simple.
{\it Phys. Rev. Lett.} {\bf 1996}, {\it 77}, 3865-3868.

\bibitem{dftu1}
Dudarev, S.L.; Botton, G.A.; Savrasov, S.Y.; Humphreys, C.J.; Sutton, A.P.
Electron-Energy -Loss Spectra and the Structural Stability of Nickel Oxide:
an LSDA+U Study.  {\it Phys.~Rev.~B} {\bf 1998}, {\it 57}, 1505-1509.

\bibitem{zhou}
Zhou, F.; Cococcioni, M.; Marianetti, C.A.; Morgan, D.; Ceder, G.
First-Principles Predictions of Redox Potentials in Transition Metal
Compounds with LDA+U.  {\it Phys. Rev. B} {\bf 2004}, {\it 70}, 235121.

\bibitem{liverpool1}
Other DFT+U implementations have been applied to spinel LMO; see
Ref.~\onlinecite{liverpool}.

\bibitem{ec}
Leung, K.; Budzien, J.L.  
{\it Ab initio} Molecular Dynamics Simulations of the Initial
Stages of Solid-electrolyte Interphase Formation on Lithium Ion Battery
Graphitic Anodes. {\it Phys. Chem. Chem. Phys.} {\bf 2010}, {\it 12}, 6583-6586.

\bibitem{yueqi}
Shi, Q.; Lu, P.; Liu, Z.; Qi, Y.; Hector Jr.; L.G.; Li, H.; Harris, S.J.
Direct Calculation of Li-ion Transport in the Solid Electrolyte Interphase.
{\it J.~Am.~Chem.~Soc.} {\bf 2012}, {\it 134}, 15476-15487.

\bibitem{pbe0}
Adamo, C.; Barone, V.  Towards Reliable Density Functional Methods without
Adjustable Parameters: the PBE0 Model.
{\it J.~Chem.~Phys.} {\bf 1999}, {\it 110}, 6158-6170.

\bibitem{sd1}
Ouyang, C.Y.; Shi, S.Q.; Wang, Z.X.; Li, H.; Huang, X.J.; Chen, L.Q.
Ab initio Molecular-dynamics Studies on Li$_x$Mn$_2$O$_4$ as Cathode Material
for Lithium Secondary Batteries.  {\it Europhys.~Lett.} {\bf 2004}, {\it 67}, 
28-34.

\bibitem{sd2}
Bo, X; Meng, Y.S.  Factors Affecting Li Mobility in Spinel LiMn$_2$O4$_4$
-- A First-Principles Study by GGA and GGA Plus U Methods.
{\it J.~Power~Sources}  {\bf 2010}, {\it 195}, 4791-4796.

\bibitem{sd3}
Islam, M.S.; Fisher, C.A.J.
Lithium and Sodium Battery Cathode Materials: Computational Insights into
Voltage, Diffusion and Nanostructural Properties.
{\it Chem.~Soc.~Rev.} {\it 2014}, {\it 43}, 185-204.

\bibitem{mo1}
Zhu, Y.; He, X.; Mo, Y.
Origin of Outstanding Stability in Lithium Solid Electrolyte Materials:
Insights from Thermodynamic Analyses Based on First-Principles Calculations.
{\it ACS Appl.~Mater.~Interfaces} {\bf 2015}, 7, 23685-23693.

\bibitem{mo2}
Mo, Y.; Ong, S.P.; Ceder, G.
First Principles Study of the Li$_{10}$GeP$_2$S$_{12}$ Lithium Super Ionic
Conductor Material.  {\it Chem.~Commun.} {\bf 2012}, {\it 24}, 15-17.

\bibitem{note}
The reaction $ 3{\rm EC} + {\rm LiMn}_2{\rm O}_4 \rightarrow 9 {\rm CO}_2 +
6 {\rm H}_2{\rm O} + 2 {\rm MnO}  + 5 {\rm Li}_2{\rm O}$
is exothermic by 0.89~eV per EC molecule at T=0~K.  This value is obtained
using the DFT/PBE0 functional.\cite{pbe0} The CO$_2$ may subsequently
combine with Li$_2$O to form Li$_2$CO$_3$.

\bibitem{shkrob16}
Shkrob, I.A.; Abraham, D.P.
Electrocatalysis Paradigm for Protection of Cathode Materials in
High-Voltage Lithium-Ion batteries.
{\it J.~Phys.~Chem.~C} {\bf 2016}, {\it 120}, 15119-15128.

\bibitem{selloni}
He, Y.B.; Tilocca, A.; Dulub, O.; Selloni, A.; Diebold, U.
Local Ordering and Electronic Signatures of Submonolayer Water on
Anatase TiO$_2$ (101).  {\it Nat.~Mater.} {\bf 2009}, {\it 8}, 585-589.

\bibitem{tio2}
Lotfabad, E.M.; Kalisvaart, P.; Cui, K.; Kohandehghan, A.; Kupsta, M.;
Olsen, B.; Mitlin, D.  ALD TiO$_2$ Coated Silicon Nanowires for Lithium
Ion Battery Anodes with Enhanced Cycling Stability and Coulombic Efficiency.
{\it J.~Mater.~A} {\bf 2013}, {\it 15}, 13646-13657.

\bibitem{dudney13}
Baggetto, L.; Dudney, N.J.; Veith, G.M.
Surface Chemistry of Metal Oxide Coated Lithium Manganese Nickel Oxide
Thin Film Cathodes Studied by XPS.
{\it Electrochim.~Acta} {\bf 2013}, {\it 90}, 135-147.

\bibitem{mgf2}
Kraytsberg, A.; Drezner, .; Auinat, M.; Shapira, A.; Solomatin N.; Axmann, P.;
Wolfart-Merens, M.; Ein-Eli, Y.  
Atomic Layer Depsition of a Particularized Protective MgF$_2$ film on a Li-Ion
Battery LiMn$_{1.5}$Ni$_{0.5}$O$_4$ Cathode Powder Material.
{\it ChemNanoMat}, {\bf 2015}, {\it 1} 577-585.

\bibitem{wolverton14}
Aykol, M.; Kirklin, S.; Wolverton, C.
Thermodynamic Aspects of Cathode Coatings for Lithium-Ion Batteries.
{\it Adv.~Energy~Mater.} {\bf 2014}, {\it 4}, 1400690.

\bibitem{tang}
Tang, M.; Newman, J. 
Why is the Solid-Electrolyte-Interphase Selective?  Through-Film
Ferrocenium Reduction on Highly Oriented Pyrolytic Graphite.
{\it J.~Electrochem.~Soc.} {\bf 2012}, {\it 159}, A1922-A1927.

\end{references}
\end{document}